\documentclass{IEEEtran}

\usepackage{graphicx}  

\usepackage{epsfig}

\usepackage{subfig} 
\usepackage{amssymb}
\usepackage{mathtools}

\usepackage{amsmath}   
\begin{document}
%
\title{Stability Analysis for Fast Settling Switched DPLL}
\author{Pallavi~Paliwal,
        Debasattam~Pal,
        and~Shalabh~Gupta
\thanks{The authors are with the Department
of Electrical Engineering, Indian Institute of Technology Bombay, Mumbai 400076, India (e-mail: pallavi.paliwal@iitb.ac.in; debasattam@ee.iitb.ac.in; shalabh@ee.iitb.ac.in).}}

\maketitle

\begin{abstract}
 In current generation digital phase locked loop (DPLL) architectures, techniques like adaptive loop bandwidth with loop order switching and switched phase-detection are employed to achieve better lock time and jitter performance. This work derives stability conditions for such DPLL architectures using Multiple Lyapunov Functions (MLFs) for switched systems. The loop-parameters chosen on the basis of these stability conditions ensure that chattering phenomenon does not occur during switching between different subsystems.  A 5\,GHz fractional-N DPLL designed with these loop-parameter values is fabricated in CMOS65\,nm-LL technology. The measured settling time of the implemented DPLL is within 1\,$\mathrm{\mu}$s. The  efficiency of switching rule and stability conditions used for this DPLL is validated with the fast settling response, which is the best lock time reported until now for fractional-N DPLLs.\\
\end{abstract}


\IEEEpeerreviewmaketitle

\section{Introduction}
In PLLs employing LC oscillators, a trade-off between settling time and output jitter exists based on the loop bandwidth. For reducing both lock-time and jitter simultaneously, the concept of adaptive loop gain has been widely used in digital phase-locked loop (DPLL) architectures \cite{ref:adaptive_pll1,ref:proposed_pll,ref:bb_deriv, ref:ml_bbpd}. Few DPLL designs also employ switching between linear and non-linear phase detection mechanism, to reduce high-resolution requirement on succeeding Time-to-Digital Converter (TDC) block. While design methodologies like adaptive gain mechanism and hybrid phase detection have been extensively explored for DPLLs, the stability analysis of these architectures has largely remained unexplored in literature. 

For PLLs involving switched subsystems, linear and steady-state s-domain analysis is unable to predict nonlinear acquisition trajectory of a system. On the other hand, analysis techniques using difference equations and state-space model can define the functioning of both linear and non-linear PLL subsystems \cite{ref:pll_linear_analysis}. This work uses Lyapunov theory for analyzing stability conditions in switched-system DPLL.

In the available literature e.g. {\cite{ref:stability_analysis_nth_power,ref:lyapunov_ref2, ref:lyapunov_ref3,ref:lyapunov_ref5, ref:lyapunov_ref6}, the stability analysis using Lyapunov theorem has been done only for PLLs involving a single Linear-Time Invariant (LTI) system. This paper investigates global asymptotic  stabilization of a  hybrid DPLL architecture \cite{ref:proposed_pll} (shown in Fig. \ref{fig:pll_block_diag}), wherein the  switching algorithm allows unstable states of non-linear time varying (NLTV) subsystem to work in coherence with linear-time invariant (LTI) subsystem. An exhaustive stability analysis becomes mandatory in this case with the architecture activating an unstable intermediate state in a bid to achieve  a record-fast response time. As part of developing the stability criteria, this work verifies the loop parameter conditions to avoid chattering phenomenon while switching between different PLL subsystems. A similar analysis framework could be  applied to other DPLL architectures as well.

As a case study, this work illustrates the  stability analysis for adaptive DPLL with loop order switching as described in Section \ref{sec:dpll_overview}. Section \ref{sec:lyapunov_overview} gives an overview of the Lyapunov theorem approach for proving stability of the DPLL system with predefined switching rule. Sections \ref{sec:lti_stability_analysis}-\ref{sec:bbpd_stability} derive Lyapunov functions for verifying the  stability of DPLL with linear phase frequency detector (PFD) and non-linear bang-bang phase detector (BBPD). Section \ref{sec:fsm_stability} outlines the stability analysis of proportional-integral-derivative (PID) controller based switching algorithm in the DPLL, based on its known region of activation and deterministic state trajectory. Section \ref{sec:sys_stability} highlights the overall DPLL stability with Lyapunov functions defined for individual subsystems. The measurement results in Section \ref{sec:meas_result} provide the proof of DPLL stability with the loop gain conditions derived in this work.


\section{DPLL System Overview}
\label{sec:dpll_overview}
We have proposed a fast settling DPLL in \cite{ref:proposed_pll} which employs variable phase-detection and adaptive loop gain to achieve optimum lock\ time performance. This architecture involves switching between different subsystems based on the magnitude of phase error. Figure \ref{fig:phase_err_fsm} shows that when the magnitude of phase error ($\phi_{err}$) is large, a linear PFD with digitally controlled oscillator (DCO) based counter is activated as the phase detection block (\textit{LTI-1 subsystem}). With reduction in the phase-error below a single DCO clock period ($\phi_{err\_1}$), an inverter based TDC is activated in the loop (\textit{LTI-2 subsystem}). When the phase-error reduces below the delay corresponding to a single inverter ($\phi_{err\_2}$), a BBPD  is activated in the loop to keep the output jitter independent of the TDC resolution. Figure \ref{fig:fsm_diag}(a) highlights the phase-error state dependent switching in the DPLL across different subsystems.

\begin{figure}[!h]
\centerline{\includegraphics[scale = 0.5]{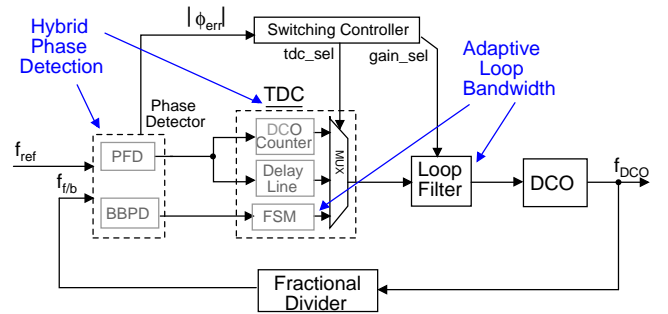}}
\caption{Detailed block diagram of Hybrid DPLL architecture.} 
\label{fig:pll_block_diag}
\end{figure}

\begin{figure}[!h]
\centerline{\includegraphics[scale = 0.65]{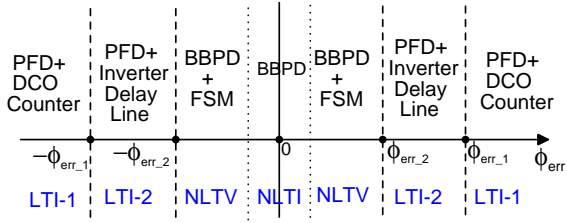}}
\caption{Phase-error state dependent switching rule for DPLL toggling between different subsystems.}
\label{fig:phase_err_fsm}
\end{figure}

To improve the settling time with bang-bang phase detection, a finite state machine (FSM) shown in Fig. \ref{fig:fsm_diag} is activated initially. This FSM emulates an additional PID controller in the loop. At the phase-error sign reversal, the FSM activates derivative gain for immediate phase correction. For a similar phase-error sign in consecutive cycles, the FSM activates another integrator ($K_{I\_{FSM}}$) in the loop to achieve fast frequency tracking. The derivative gain ($K_D$) is reduced with each phase error sign reversal. When the derivative gain becomes 0, the FSM is removed from the loop to avoid chattering in the settled state. Therefore, the BBPD+FSM emulates a Non-Linear Time Varying (NLTV) subsystem, and the BBPD without FSM represents a Non-Linear Time Invariant (NLTI) subsystem.
\begin{figure}[!h]
{
\subfloat[]{\includegraphics[scale = 0.36]{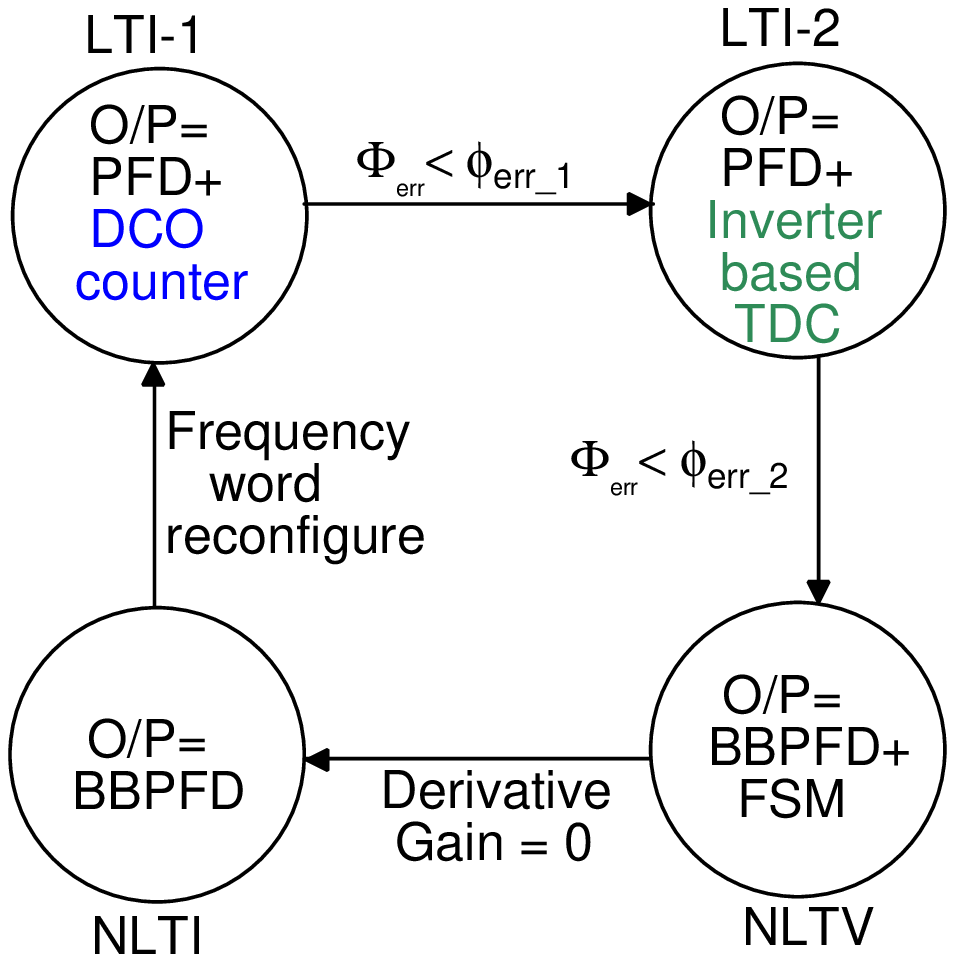}}
\subfloat[]{\includegraphics[scale = 0.37]{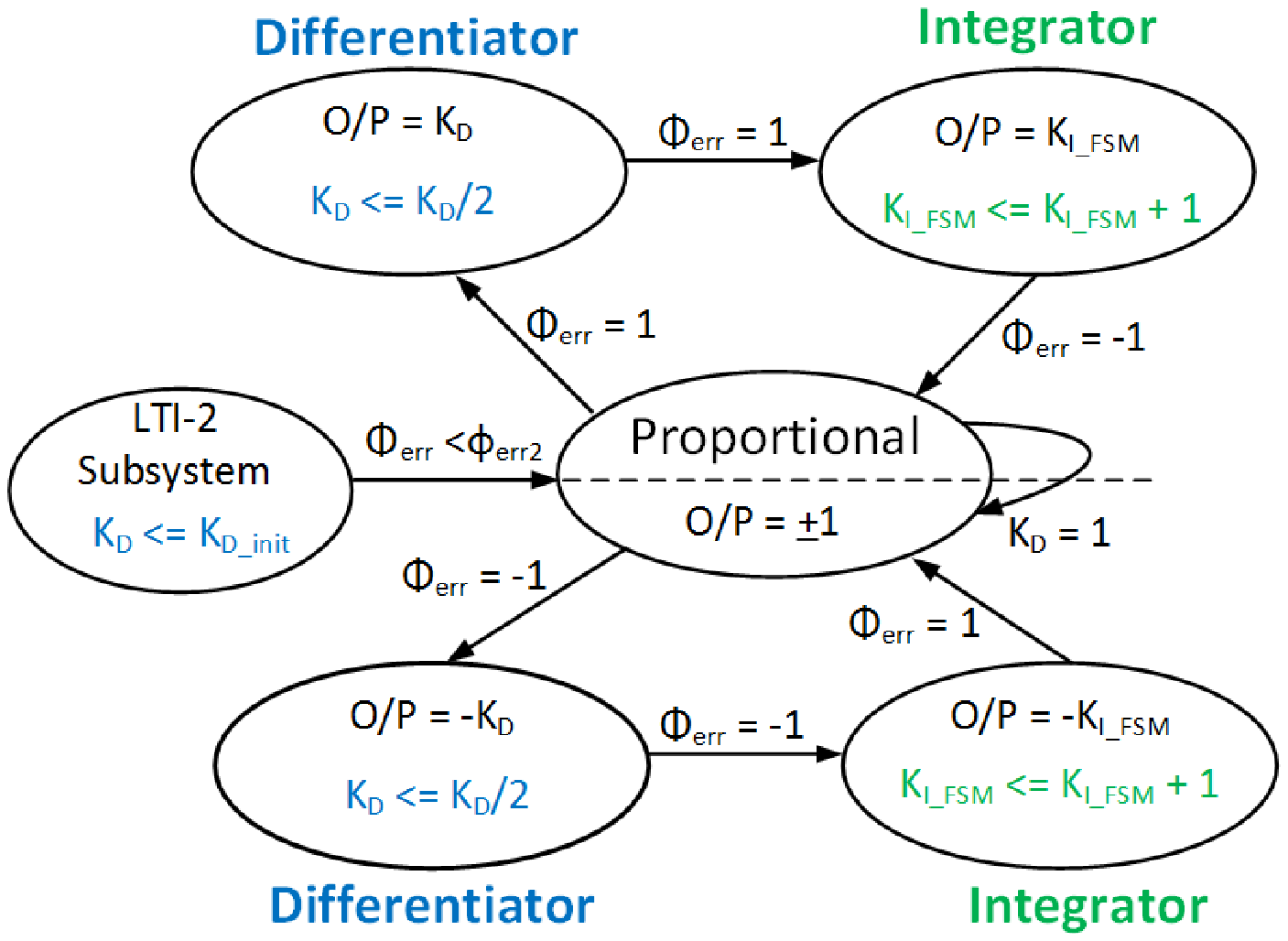}}
}
\caption{ (a) PFD+TDC variant activation based on the input phase error magnitude; (b) PID controller based FSM algorithm activated with BBPD mode.}
\label{fig:fsm_diag}
\end{figure}

The switched DPLL system is implemented in CMOS65\,nm-LL process for its performance validation, with the design parameters of the individual blocks being as stated in Table \ref{tab:design_parameters}. The phase detector characteristics in Fig. \ref{fig:bbpd_tdc_io}(a)  highlight that the loop switching allows to overcome the limitation of finite resolution of DCO based Counter and delay line functioning as the TDC block. The impulsive response in BBPD+FSM characteristic in Fig. \ref{fig:bbpd_tdc_io}(b), along with additional accumulator, allows rapid phase and frequency error correction. To summarize, the DPLL achieves fast lock and low jitter with its phase margin and loop bandwidth varied across the switched-subsystem (Fig. \ref{fig:phase_err_fsm}) as shown in Table \ref{tab:loop_parameters}.
  
\begin{table}[!h]
 \caption{DPLL Circuit Parameters}
\label{tab:design_parameters}
\begin{center}
\renewcommand{\arraystretch}{1.4}
  \begin{tabular}{ | c | c | c |}
    \hline
 Output Frequency & 4.8\,GHz- 5.02\,GHz \\  
 Reference Clock ($f_{ref}$) &  100\,MHz\\
 DCO Gain ($K_{DCO}$) &   10\,KHz/LSB \\ 
 DCO Counter Resolution ($\phi_{err\_1}$) &  1.67\,ns \\
 Inverter Delay Line Resolution ($\phi_{err\_2}$) & 20\,ps  \\
 DCO Jitter ($\sigma_{t_{DCO}}$) & 0.2\,ps \\ \hline
    \end{tabular}
\end{center}
\end{table}

\begin{figure}[!h]
\centering
{
\subfloat[]{
{\includegraphics[scale = 0.45]{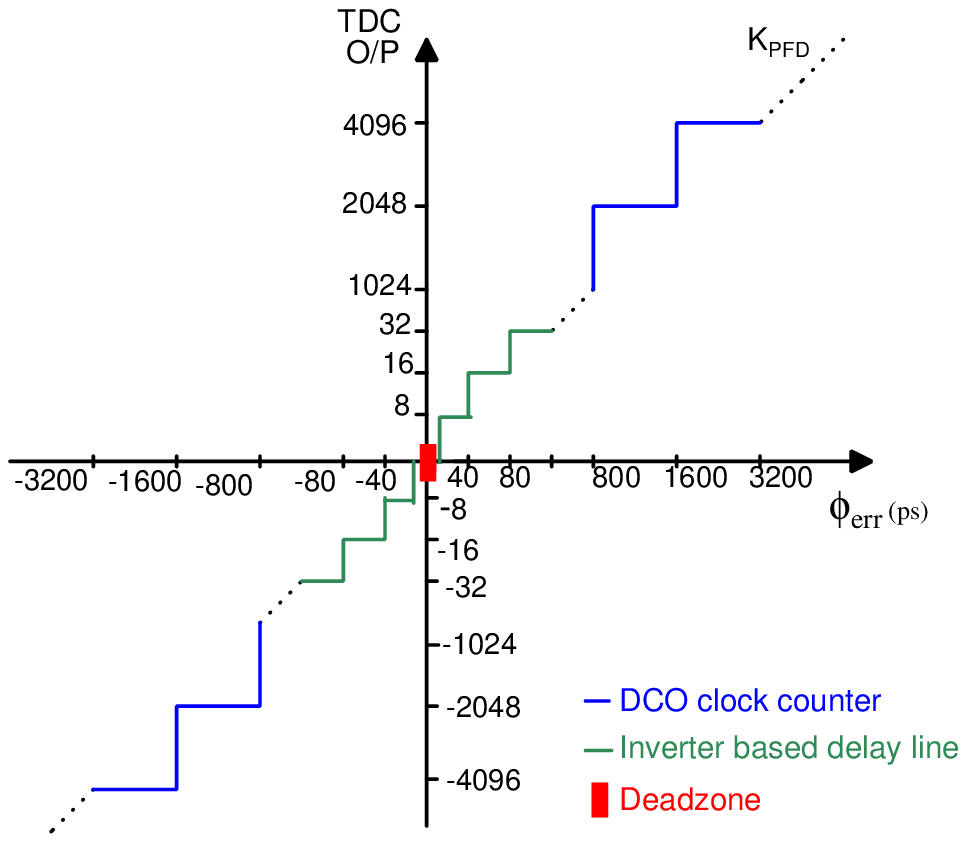}}}
\subfloat[]{
{\includegraphics[scale = 0.35]{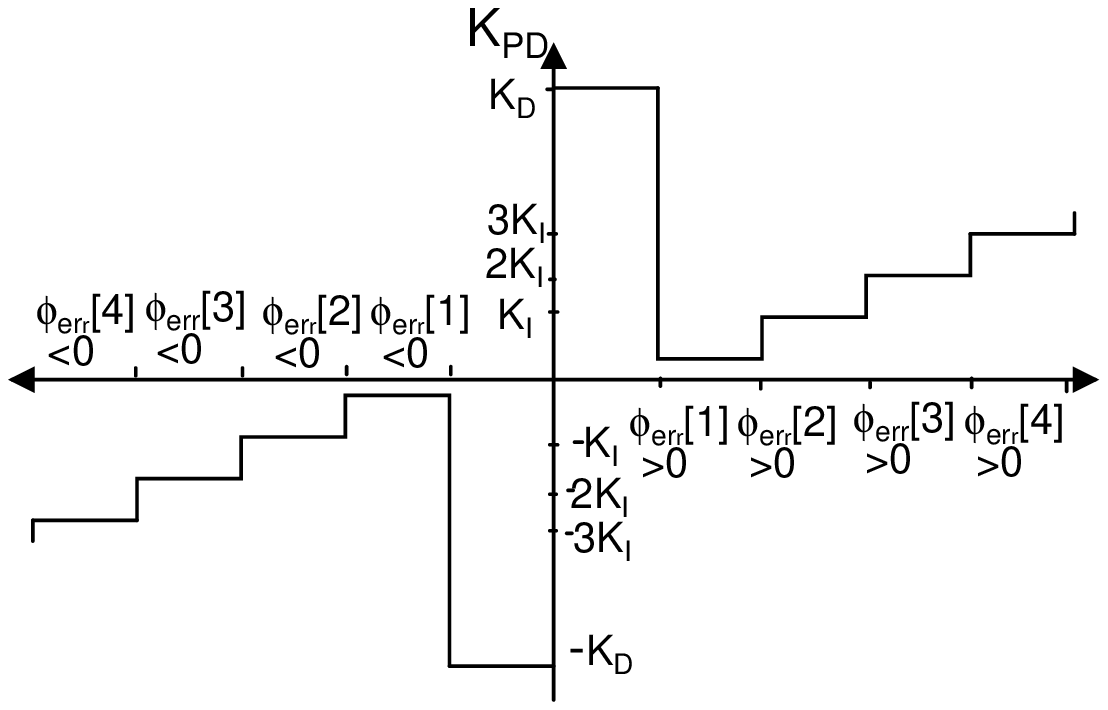}}}
}
\caption{(a)  I/O characteristics for  PFD activated with  DCO Counter for $|\phi_{err}|>|\phi_{err\_1}|$, and inverter delay line for $|\phi_{err}|>|\phi_{err\_2}|$; (b) I/O characteristics for  BBPD+FSM.}
\label{fig:bbpd_tdc_io}
\end{figure}

\begin{table}[!h]
 \caption{Switched DPLL System Parameters}
\label{tab:loop_parameters}
\begin{center}
\renewcommand{\arraystretch}{1.4}
  \begin{tabular}{  c  c c  c }
    \hline
    \textbf{Phase Error}  & \textbf{Phase} & \textbf{Loop Type} & \textbf{Parameters}\\ 
    \textbf{State}  &   \textbf{Detection} & & \\ \hline
$>\phi_{err\_1}$ & PFD+ & LTI with  & PM  $\approx15^\circ$ \\
                         &  DCO Counter & high loop gain & B/W = 10\,MHz \\ \hline
$>\phi_{err\_2}$ & PFD+ & LTI with  & PM  $\approx35^\circ$ \\
                         &  Delay Line & moderate loop gain & B/W = 7\,MHz \\ \hline                  
 $<\phi_{err\_2}$ & BBPD+ & NLTV  & Unstable \\
  $(K_D > 0)$        &  FSM &  &  \\ \hline   
  $<\phi_{err\_2}$ & BBPD & NLTI with & B/W = 2\,MHz \\
  $(K_D = 0)$        &  &low loop gain  &  \\ \hline    
    \end{tabular}
\end{center}
\end{table}


\section{Lyapunov Theorem overview}
\label{sec:lyapunov_overview}
Lyapunov theorem states that if a positive-definite storage function of the system could be found, which is  decreasing with time along every possible trajectory of the system, then the equilibrium point is asymptotically stable. Thus, the system's stability in Lyapunov sense is stated based on the storage function as follows: 

Consider a discrete system described by $x_{k+1} = f(x_k)$, where $f(0) = 0$, $x_k \in R^n$, $k$ being the time index. The equilibrium state $x = 0$ is globally asymptotically stable if there exists a continuous scalar function $V_k = V(x_k)$, such that, 

\qquad $1.\ V(0) = 0$

\qquad $2.\ V(x_k) > 0 \ \forall \  x\neq0 $

\qquad $3.\ \lim_{x_k\rightarrow\infty}V(x_k)\ =\  \infty$

\qquad $4.\ \triangle{V_k}\equiv V(x_{k+1}) - V(x_k) < 0 \ \forall \  x\neq0$

\noindent Hence for a stable system, the design parameters are chosen such that the candidate Lyapunov function ($V(x_k)$) has a negative derivative along the trajectory of the system. 

The DPLL in Fig. \ref{fig:pll_block_diag} is a state-dependent switched system. Though the subsystems of DPLL could be individually stable, the loop may still have divergent trajectories.
On the other hand, it is also possible to make the loop switch between unstable subsystems such that the complete switched system becomes asymptotically stable \cite{book:switched_system}. Therefore, based on stability conditions, the DPLL parameters are derived such that the loop converges to its equilibrium point without cycling infinitely between two or more subsystems.

In a switched system, if there exists a Common Quadratic Lyapunov Function (CQLF) for certain subsystems, then the system is stable under arbitrary switching between those subsystems. Section \ref{sec:lti_stability_analysis} discusses the loop parameter derivation based on the CQLF condition for stability, while switching between LTI-1 and LTI-2 subsystem.

The stability analysis of nonlinear-system is facilitated by the fact that Lyapunov function derivative only needs to be evaluated in the region where the sytem would be active \cite{book:switched_system}. Based on this clause, Section \ref{sec:bbpd_stability} discusses the stability analysis of BBPD based DPLL, wherein a nonlinear function is replaced by different values it assumes in various regions of the phase plane. The state-space is thus partitioned into different regions based on the varying values of non-linear function. Consequently, the stability condition as per Lyapunov function is derived for these different cases of state-space equations. 

With a suitable switching law in a switched subsystem, the trajectory can be brought to equilibrium even in  the presence of an unstable subsystem. To achieve this, the stable subsystems are activated to compensate the state divergence caused by an unstable subsystem. Following this concept, Section-\ref{sec:fsm_stability} verifies the stability of BBPD+FSM based DPLL which incorporates an unstable third-order integrator based subsystem. In this case, the energy functions of subsystems are analyzed at switching instants instead of consecutive cycles. Mutiple Lyapunov Functions method \cite{ref:mult_lyapunov_stability} states that for stability of system with difference equation $x_{k+1}=f_j(x_{k})$, the candidate Lyapunov function $V_j$ corresponding to $j$th subsystem should satisfy condition in (\ref{eqn:mult_lyapunov}), as shown in Fig. \ref{fig:mult_lyapunov_stability}.
  \begin{equation}
\label{eqn:mult_lyapunov}
V_j(x_j(\overline{t}_{j,k+1}))<V_j(x_j(\overline{t}_{j,k}))
\end{equation} 

\noindent where, ${\overline{t}_{j,k}},{\overline{t}_{j,k+1}}$ are two consecutive switch-on instants of subsystem j.


\begin{figure}[!h]
\centerline{\includegraphics[scale=0.65]{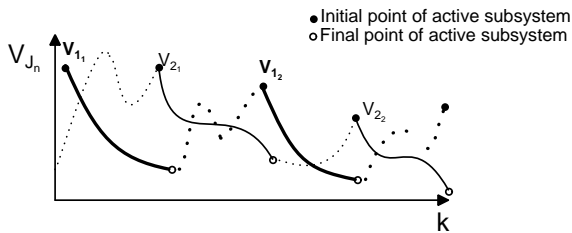}}
\caption{Stability with multiple-Lyapunov functions is decided with decreasing value of $V_j$, only at switching instants \cite{ref:mult_lyapunov_stability}.}
\label{fig:mult_lyapunov_stability}
\end{figure}
Thus, the multiple Lyapunov functions approach could be used to show stability for overall loop of DPLL incorporating subsystems with different storage functions.

\section{Linear DPLL stability analysis}
\label{sec:lti_stability_analysis}
Consider that LTI-1/LTI-2 subsystem of DPLL, represented by Fig. \ref{fig:pll_model}, have initial phase and frequency error of $\phi_0$ and $\Delta{f_0}$. For evaluating PLL behaviour as an autonomous system, consider $\phi_k$ and $\Delta\phi_{f\_{k}}$ as state variables representing phase error and additive phase due to frequency error in $k^{th}$ sampling instant. 
\begin{figure}[!h]
\centerline{\includegraphics[scale=0.65]{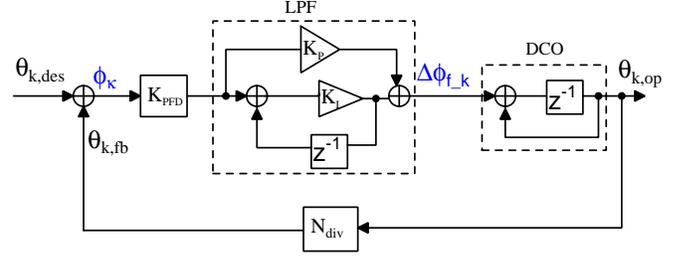}}
\caption{Discrete-time model of second order PLL}
\label{fig:pll_model}
\end{figure}

\noindent With an ideal reference clock ($f_{ref}$) and a constant desired phase ($\theta_{k,des}$), the state-space equations for LTI-1/LTI-2 subsystems of DPLL are given as 
 \begin{equation}
\label{eqn:phi_lin2}
\begin{aligned}
& \phi_{k+1}=\phi_{k}+{\Delta\phi_{f\_k}} - K_P^\prime(\phi_{k}-\phi_{k-1}) - K_I^\prime\phi_k,\\
& \Delta\phi_{f\_{k+1}}=\Delta\phi_{f\_k} - K_P^\prime(\phi_{k}-\phi_{k-1}) - K_I^\prime\phi_k,
\end{aligned}
\end{equation} 
\noindent where  $\phi_{k} = \theta_{k,des} - \theta_{k,fb}$, with $\theta_{k,fb}$ being the feedback phase in the loop. The variables $K_P^\prime$ and $K_I^\prime$ denoting the proportional and integral gain in the loop, with $K_{PFD}$ being the TDC gain, are calculated as 
 \begin{equation}
\label{eqn:loop_gain_var}
\begin{aligned}
K_P^\prime &= \frac{K_{PFD}K_P2\pi{K_{DCO}}}{N_{div}f_{ref}},\\
K_I^\prime &= \frac{K_{PFD}K_I2\pi{K_{DCO}}}{N_{div}f_{ref}},\\
&K_{PFD} = \frac{T_{ref}}{2\pi\Delta{t_{TDC}}},
\end{aligned}
\end{equation} 

\noindent where  $K_{DCO}$ is the DCO gain, $N_{div}$ is the feedback division ratio, $\Delta{t_{TDC}}$ is the TDC resolution, and $T_{ref}$ is the reference clock period.

During locking process in PLL, the phase error ($\phi_{k}$) is corrected by modulating the DCO frequency. Hence, the phase-error derivative is related to the incremental phase ($\Delta\phi_{f\_k}$) from frequency error at the PLL output as 
 \begin{equation}
\label{eqn:ph_freq_rel}
{\Delta\phi_{f\_k}} = \phi_{k}-\phi_{k-1}.
\end{equation} 

The PLL is settled when the system's trajectory reaches an equilibrium point of phase and frequency error ($\phi_k,\Delta\phi_{f\_k}$) being (0,0). From (\ref{eqn:phi_lin2})-(\ref{eqn:ph_freq_rel}), the state space equations used to prove stability in Lyapunov sense are derived as 

\begin{equation}
\label{sseqn:phi_lin2}
\begin{aligned}
& \phi_{k+1}= (1-K_I^\prime)\phi_k+ (1 - K_P^\prime)\Delta\phi_{f\_{k}},  \\
& \Delta\phi_{f\_{k+1}}= -K_I^\prime\phi_k +(1 - K_P^\prime)\Delta\phi_{f\_{k}}.  
\end{aligned}
\end{equation} 

\noindent Representing the state-space equations in the form of matrices (\ref{eqn:sys_matrix}), the system equation can be written as (\ref{eqn:sys_eqn}).

\begin{equation}
\label{eqn:sys_matrix}
\begin{bmatrix} {\phi}_{k+1} \\ \Delta\phi_{f\_{k+1}} \end{bmatrix} = \left[ \begin{array}{cc} 1-K_I^\prime & K   \\-K_I^\prime & K \end{array} \right]  \left[ \begin{array}{c}  {\phi}_{k} \\ \Delta\phi_{f\_{k}} \end{array} \right] 
\end{equation}

\noindent where  $K = 1 - K_P^\prime$

\begin{equation}
\label{eqn:sys_eqn}
{x}_{k+1}=Ax_k
\end{equation}

\noindent For stability analysis of this system,  the candidate quadratic Lyapunov function is chosen as \\
 \begin{equation}
\label{eqn:lyapunov_eqn1}
V_{k}=x^T_kPx_k,
 \end{equation}
\noindent where  $P= \left[ \begin{array}{cc} p_{11} & p_{12}\\ p_{12} & p_{22}\end{array}\right]$. \\

\noindent To obtain $V_k$ as a positive quantity, matrix P should be positive definite, which leads to the condition:

\begin{center}
$p_{11}p_{22} > p_{12}^2$, $p_{11}>0$, $p_{22}>0$.
\end{center}

\noindent Using the Lyapunov function $V_k$, the change in energy of the system is given by 
\begin{equation}
\label{eqn:lyapunov_linear}
{\Delta}V_k = x^T_k(A^TPA-P)x_k.
\end{equation}

\noindent For the system to be stable in Lyapunov sense, $(A^TPA-P)$ must be negative definite to have ${\Delta}V_k < 0 \ \forall \  x_k\neq0$. Here, the discrete-time Lyapunov equation is $A^TPA-P+Q=0$ with $Q>0$, wherein for $A^TPA$ being the energy function, $A^TQA$ could be considered the associated dissipation. 

Based on the required phase-margin and unity-gain bandwidth in the considered DPLL, loop-filter gain ($K_{P_i}/K_{I_i}$) for LTI-1 and LTI-2 subsystem is calculated using linear z-domain analysis presented in \cite{ref:pfd_analysis}\cite{ref:proposed_pll} as 
\begin{equation}
\label{eqn:kp_val}
\begin{aligned}
&K_P = \frac{N}{K_{PFD}K_{DCO}}.\frac{\omega_{UGBW}}{\sqrt{1+{tan^{-2}(PM)}}}\left(1 - \frac{T_{ref}}{2}.\omega_z\right), \\
&K_I = T_{ref}\frac{N}{K_{PFD}K_{DCO}}.\frac{\omega^2_{UGBW}}{\sqrt{1+{tan^2(PM)}}},\\
\end{aligned}
\end{equation}
\noindent where $T_{ref}$ is the reference clock period, $N_{div}$ is the feedback divider, ${\omega_{UGBW}}$ is the unity-gain bandwidth and $PM$ is the phase-margin of the loop. Figure \ref{fig:lti_1_phase_portrait_pll} shows the converging trajectory of DPLL in LTI-1 and LTI-2 mode, with loop-filter gain in Table \ref{tab:linear_filter_gain} being derived from (\ref{eqn:kp_val}), based on the loop-bandwidth and phase-margin desired in each mode.

\begin{table}[!h]
 \caption{Filter Gain with Linear Phase Detection}
\label{tab:linear_filter_gain}
\begin{center}
\renewcommand{\arraystretch}{1.4}
  \begin{tabular}{ | c | c | c |}
    \hline
\textbf{Gain} &  \textbf{LTI-1 subsystem}  & \textbf{LTI-2 subsystem}  \\ \hline
 $K_{PFD}$ & ${6}/{2\pi}$&  ${250}/{2\pi}$ \\  
 \textit{(LSB/rad)} & & \\
$K_P$  & 4096 & 128  \\ 
$K_I$  & 1024 &  8 \\ \hline
    \end{tabular}
\end{center}
\end{table}

\begin{figure}[!h]
{
\subfloat[]{
\includegraphics[scale=0.17]{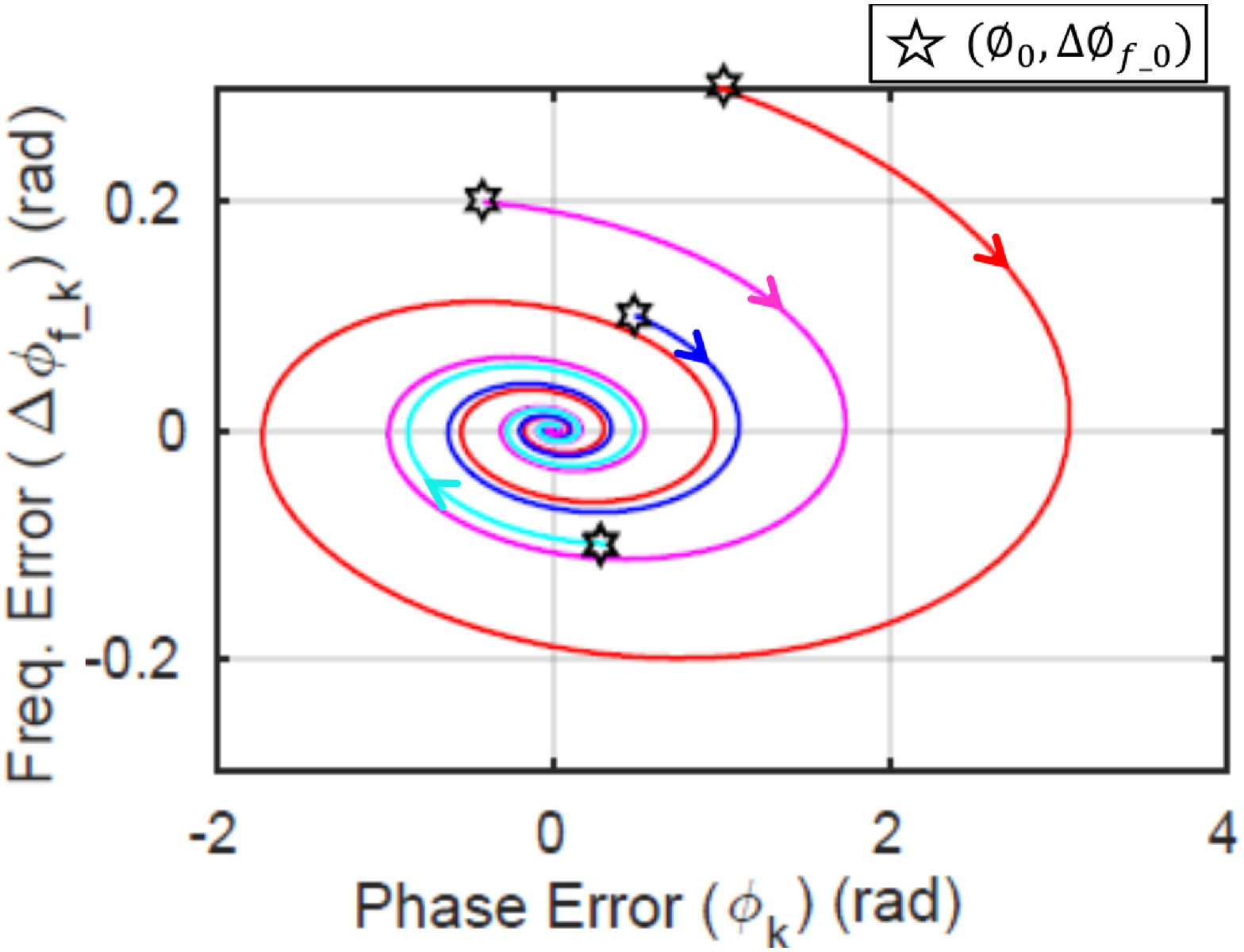}}
\subfloat[]{
\includegraphics[scale=0.16]{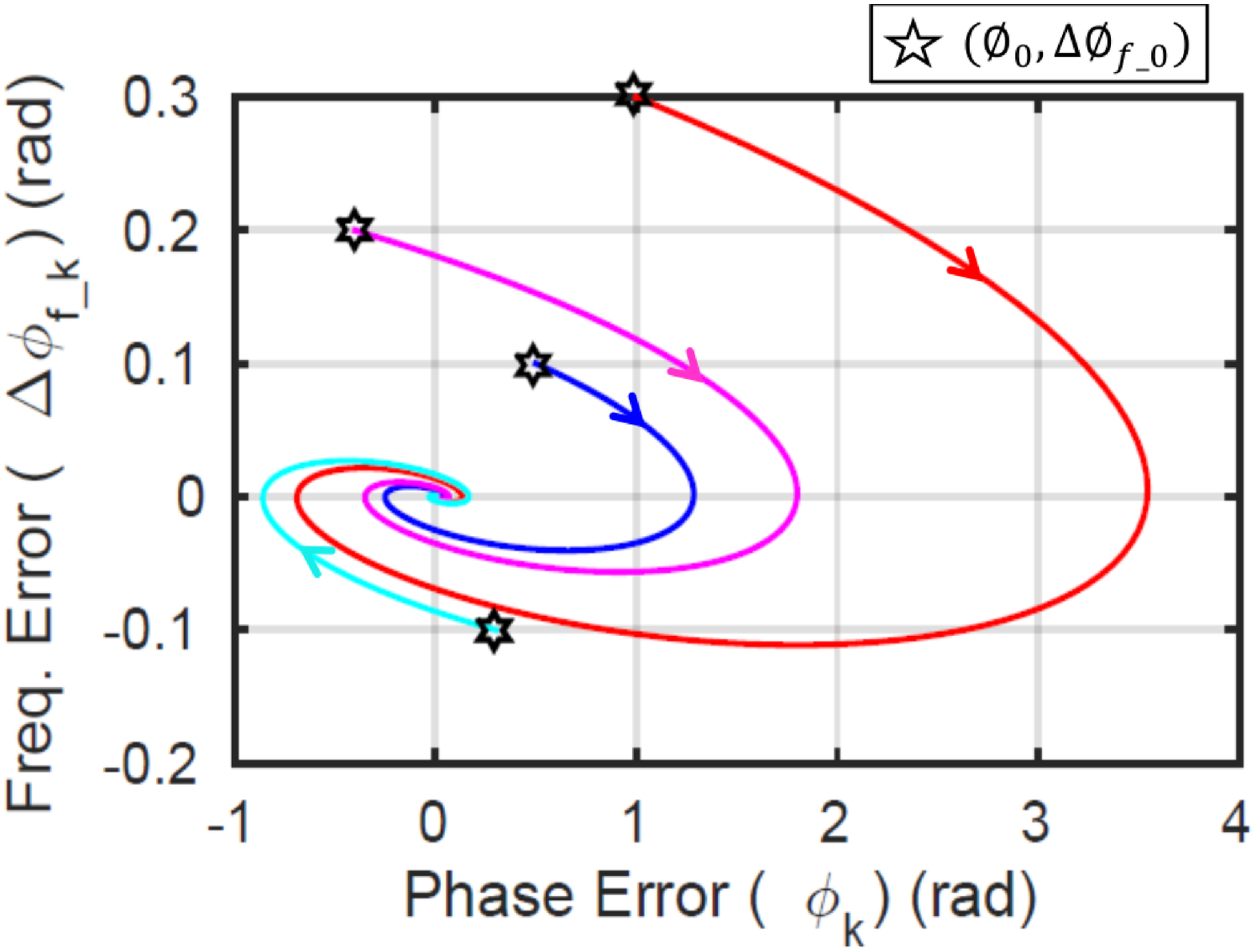}}
}
\caption{Phase-portrait of PLL in LTI-1 and LTI-2 mode with  different initial conditions $(\phi_0,\Delta\phi_{f\_0})$ in the phase-plane. Loop-filter gain in (a) LTI-1 mode: $K_{P1}'=0.03\mathrm{rad}$, $K_{I1}'=0.007\mathrm{rad}$, and (b) LTI-2 mode: $K_{P2}'=0.05\mathrm{rad}$, $K_{I2}'=0.003\mathrm{rad}$.}
\label{fig:lti_1_phase_portrait_pll}
\end{figure}

Corresponding to the gain ($K_{P_i}$,$K_{I_i}$) in each subsystem, the matrix P could be found for common Lyapunov Function to have decreasing derivative under arbitrary switching between LTI-1 and LTI-2 subsystem. For instance, Fig. \ref{fig:energy_rate_lti_pll} shows the settling response of DPLL  while switching between LTI-1 and LTI-2 subsystem with CQLF derived as
\begin{equation}
\label{eqn:cqlf}
 V_{1,k} =\left[\begin{array}{cc} {\phi}_{k} & \Delta\phi_{f\_{k}} \end{array} \right]\left[ \begin{array}{cc} 0.02 & 0.06\\0.06 & 3\end{array} \right]  \begin{bmatrix} {\phi}_{k} \\ \Delta\phi_{f\_{k}} \end{bmatrix}.
\end{equation}

\begin{figure}[!h]
{
\subfloat[]{
\includegraphics[scale=0.19,trim=4.5cm 0 0 1.5cm, clip]{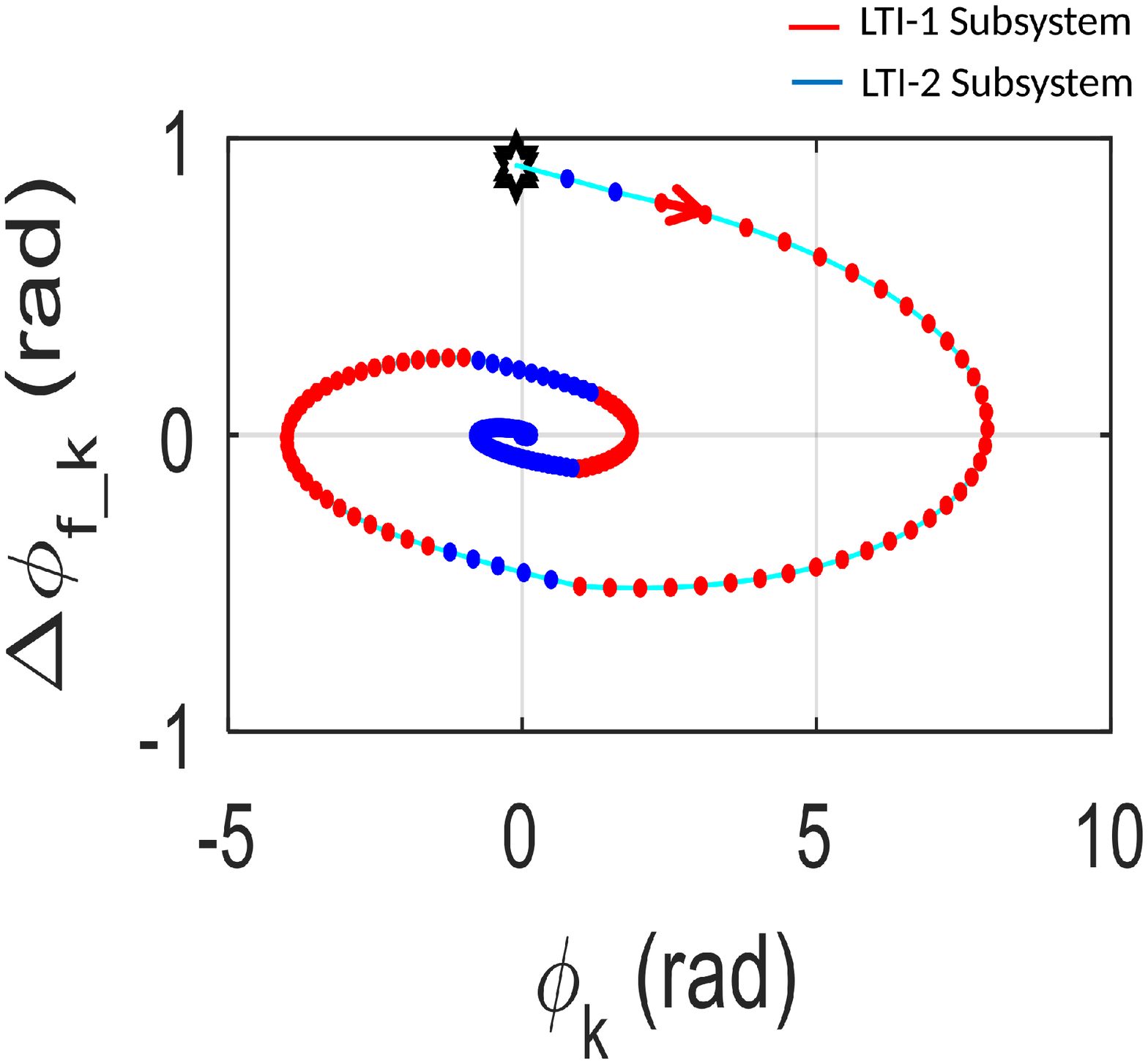}}
\subfloat[]{
\includegraphics[scale=0.19,trim=0 0 1cm 0.5cm, clip]{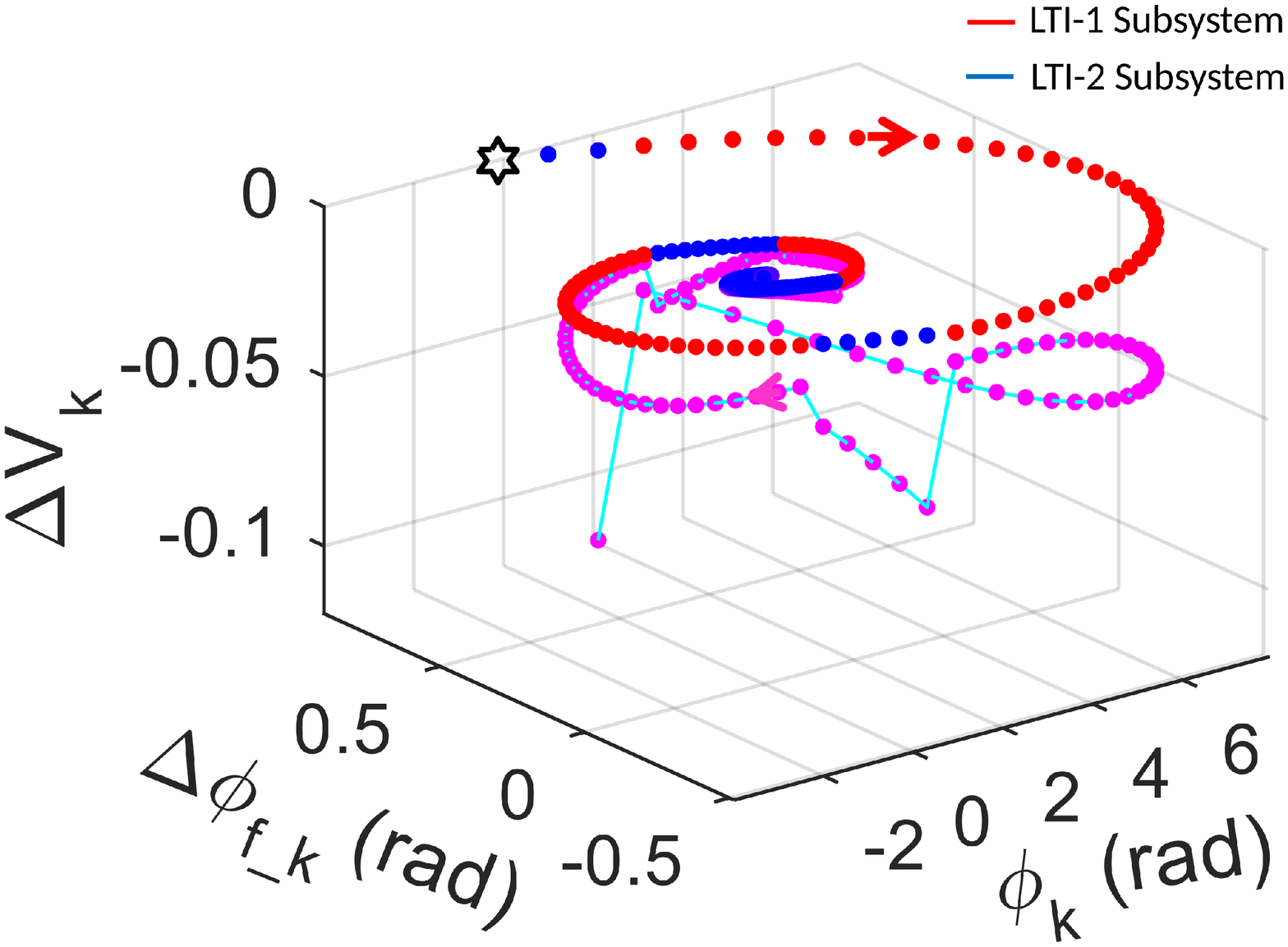}}
}
\caption{(a) System trajectory with phase-error state-dependent switching between LTI-1 and LTI-2 subsystem; (b) Negative energy-derivative curve for DPLL with projection of switched LTI-1/LTI-2 subsystems' trajectory.}
\label{fig:energy_rate_lti_pll}
\end{figure}

Though the stability conditions in individual LTI-subsystem could be derived with linear z-domian analysis, the Lyapunov theorem aids in proving that the DPLL will be stable under arbitrary switching between LTI-1 and LTI-2 subsystem.  


\section{BBPD based DPLL stability }
\label{sec:bbpd_stability}

Figure \ref{fig:bbpd_dpll_model} shows the non-linear operation of BBPD based DPLL. In this case, the phase detector output $\sigma(\phi_{k})$ takes up the binary values of either +1 or -1, based on the sign of input phase error ($\phi_{k}$). 

\begin{figure}[h]
\centerline{\includegraphics[scale=0.65]{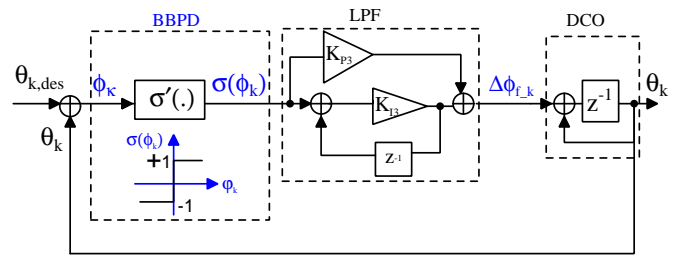}}
\caption{Discrete-time model of BBPD based DPLL.}
\label{fig:bbpd_dpll_model}
\end{figure}

\noindent The DPLL state-space equations representing BBPD operation are  
\begin{equation}
\label{eqn:phi_bbpd}
\begin{aligned}
&\phi_{k+1}=\phi_{k}+{\Delta\phi_{f\_k}} - K_{P3}'[\sigma(\phi_{k})-\sigma(\phi_{k-1})] - K_{I3}'\sigma(\phi_k),\\
&\Delta\phi_{f\_k+1}={\Delta\phi_{f\_k}} - K_{P3}'[\sigma(\phi_{k})-\sigma(\phi_{k-1})]  - K_{I3}'\sigma(\phi_k),
\end{aligned}
\end{equation} 

\noindent where $K_{P3}'$ and $K_{I3}'$ representing the proportional and integral gains in the feedforward path of BBPD based DPLL are given as

 \begin{equation}
\label{eqn:loop_gain_var}
\begin{aligned}
K_{P3}^\prime &= \frac{K_{P3}2\pi{K_{DCO}}}{N_{div}f_{ref}},\\
K_{I3}^\prime &= \frac{K_{I3}2\pi{K_{DCO}}}{N_{div}f_{ref}}.\\
\end{aligned}
\end{equation}

\subsection{State-Space Partition}

For simplified analysis, BBPD based nonlinear system can be modeled as a piecewise affine system of the form, 
\begin{equation}
\label{eqn:pwa}
x_{k+1}=A_ix_k+a_i \text{\quad \quad  for\ } x_k\in R_i
\end{equation}

\noindent where $R_i$'s are polyhedral partitions of the state-space. The DPLL state-space can be thus partitioned into regions shown in Fig. \ref{fig:bbpd_phase_plane_eqns}, with the corresponding state-space equations derived from (\ref{eqn:phi_bbpd}). Hence, based on the location of the state, the system's settling trajectory is governed by (\ref{eqn:phi_bbpd1}) or (\ref{eqn:phi_bbpd2}), depending on whether the phase-error sign reverses or remains same in consecutive cycles.
\begin{equation}
\label{eqn:phi_bbpd1}
\begin{aligned}
&\phi_{k+1}=\phi_{k}+{\Delta\phi_{f\_k}} - (2K_{P3}'+K_{I3}')\sigma(\phi_k)\\
&\Delta\phi_{f\_k+1}={\Delta\phi_{f\_k}} - (2K_{P3}'+K_{I3}')\sigma(\phi_k)
\end{aligned}
\end{equation} 
\begin{equation}
\label{eqn:phi_bbpd2}
\begin{aligned}
&\phi_{k+1}=\phi_{k}+{\Delta\phi_{f\_k}} - K_{I3}'\sigma(\phi_k)\\
&\Delta\phi_{f\_k+1}={\Delta\phi_{f\_k}} - K_{I3}'\sigma(\phi_k)
\end{aligned}
\end{equation}

\begin{figure}[!h]
\centerline{\includegraphics[scale=0.65]{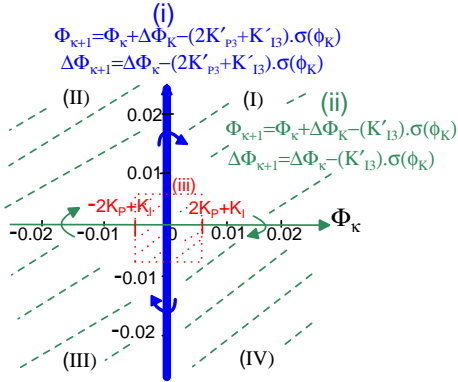}}
\caption{BBPD based DPLL phase plane with state-space equations governing DPLL settling trajectory during (i) phase-error sign reversal, (ii) similar phase-error sign in consecutive cycles, and (iii) limit cycle region.}
\label{fig:bbpd_phase_plane_eqns}
\end{figure}

\subsection{State-Trajectory Direction}
Equation (\ref{eqn:phi_bbpd}) indicates that starting from an arbitrary $(\phi_0,\Delta\phi_0)$  point, the system's trajectory moves in clockwise direction for $K_{P3}'>0$ and $K_{I3}'>0$. Also, (\ref{eqn:phi_bbpd}) shows that the DPLL system enters into limit-cycle regime for $(\phi_k,\Delta\phi_k)$ satisfying condition $\phi_k+\Delta\phi_k < (2K_{P3}'+K_{I3}')\sigma(\phi_k)$. Hence in succeeding sections, the Lyapunov function is discussed only for quadrant change by the DPLL trajectory in clockwise direction, outside the limit-cycle region. This knowledge of the state trajectory direction while passing through the partitioned state-space region aids in deriving the energy decreasing condition of the system.
%
%

\subsection{Choice of Loop Filter Gain}
Ideally, the lowest possible proportional gain (${K_{P3}'}$) results in minimum jitter from BBPD limit cycles. With the lowest proportional gain (${K_{P3}'}$) in the implemented DPLL, the DCO gain ($K_{DCO}$) of 10\,KHz/LSB corresponds to phase step ($\triangle\phi_{DCO,LSB}$) of only 15\,fs in one DCO cycle. This low magnitude of phase step lies within the DCO's random noise regime (${\sigma_{t,DCO}}\approx200\,fs$, as stated in Table \ref{tab:loop_parameters}), and does not result in an immediate phase error correction. Thus, the decision of filter-proportional gain ($K_{P3}$) value depends upon the jitter in the reference and feedback clocks. A lower proportional gain results in less noise contribution from  BBPD limit cycles. However, a  higher proportional gain ensures that a larger jitter in the feedback clock can be tolerated, without causing the DPLL to switch between different subsystems. In this design, the proportional gain ($K_{P3}$) for final settling with BBPD is approximately kept as ${\sigma_{t,DCO}}/{\triangle\phi_{DCO,LSB}}$ ($\approx 8$). Based on a fixed proportional gain ($K_{P3}$), the integral gain ($K_{I3}$) of loop filter is derived for stability as {$1$} using 
\begin{equation}
\label{eqn:gain_ratio}
\frac{K_{P3}}{K_{I3}}\geq\frac{tan(\omega_uT_{ref}D+PM)}{\omega_uT_{ref}},
\end{equation}

\noindent where $D$ refers to the loop delay \cite{ref:nonlinear_analsys}.

\subsection{Candidate Lyapunov Function}
For analyzing stability in lyapunov sense, let BBPD based DPLL's energy function be represented by (\ref{eqn:bbpd_energy}), with energy function curve as shown in Fig. \ref{fig:energy_plot_bbpdpll}.
\begin{equation}
\label{eqn:bbpd_energy}
V_{3,k}= x^T_kPx_k 
\end{equation}

\noindent where $x_k = \begin{bmatrix} {\phi}_{k} \\ \Delta\phi_{f\_{k}} \end{bmatrix}$ and $P =\left[ \begin{array}{cc} 1 & 0 \\0 & 1000\end{array} \right]$ .\\

\begin{figure}[!h]
\centerline{\includegraphics[scale=0.32]{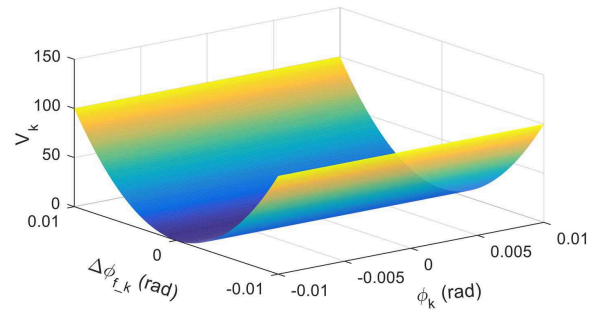}}
\caption{Candidate Lyapunov function for BBPD based DPLL.}
\label{fig:energy_plot_bbpdpll}
\end{figure}

\noindent For system $x_{k+1}=Ax_k+a$, the change in energy at each sampling instant is given by 
\begin{equation}
\label{eqn:bbpd_energy_change}
\begin{aligned}
\Delta{V_{k+1}} &= V_{k+1}-V_k,\\
&= x^T_{k+1}Px_{k+1} - x^T_kPx_k,\\ 
&= [Ax_k+a]^TP[Ax_k+a] - x^T_kPx_k,\\
&= x^T_k[A^TPA-P]x_k+2a^TPAx_k+a^TPa. 
\end{aligned}
\end{equation}

\noindent The energy change rate is evaluated for two different regions of DPLL phase plane as follows:\\

\textit{Case I: Phase-error sign reversal}\\
In this case, the state-space matrix is modified as $A = \left[ \begin{array}{cc} 1 & 1 \\0 & 1\end{array} \right]$ and $a_1 =\begin{bmatrix} -(2K_{P3}'+K_{I3}')\sigma(\phi_k) \\ -(2K_{P3}'+K_{I3}')\sigma(\phi_k) \end{bmatrix} $.\\

\noindent Accordingly, the condition on loop-filter values for decreasing energy is derived as (\ref{eqn:phase_err_sign_reverse1}), using $\Delta{V_{k+1}}$ expression from (\ref{eqn:bbpd_energy_change}).
\begin{center}
  $\Delta{V_{k+1}}<0$
\end{center}
\begin{equation}
\label{eqn:phase_err_sign_reverse1}
\begin{aligned}
\Rightarrow 1001(2K_{P3}'+K_{I3}')^2-2(\phi_k+1001\Delta\phi_{f\_k})(2K_{P3}'\\
+K_{I3}')\sigma(\phi_k)+({\phi_k^2}+{\Delta\phi_{f\_k}^2}+\phi_k\Delta\phi_{f\_k})<0.
\end{aligned}
\end{equation} \\

\textit{Case II: Similar phase error sign in consecutive cycles}\\
Here, the state space matrix is modified as $A = \left[ \begin{array}{cc} 1 & 1 \\0 & 1\end{array} \right]$ and $a_2 =\begin{bmatrix} -K_{I3}'\sigma(\phi_k) \\ -K_{I3}'\sigma(\phi_k) \end{bmatrix} $. For these values of state matrix, the energy change rate, defined in (\ref{eqn:bbpd_energy_change}), is obtained as 

\begin{equation}
\label{eqn:phase_err_sign_same}
\begin{aligned}
\Delta{V_{3,k+1}} = 1001{(K_{I3}')}^2-2(\phi_k+1001\Delta\phi_{f\_k})K_{I3}'\sigma(\phi_k)+\\
({\phi_k^2}+{\Delta\phi_{f\_k}^2}+\phi_k\Delta\phi_{f\_k}).
\end{aligned}
\end{equation} 
Equation (\ref{eqn:phase_err_sign_same}) indicates that the system's energy decreases in Quadrants (I)/(III) but increases in Quadrants (II)/(IV) of the phase plane. The increase/decrease of energy in different quadrants, is of the same order ($2002K_{I3}'|\Delta\phi_{f}| - |\phi||\Delta\phi_{f}|$) for similar magnitude of ($|\phi_f|,|\Delta\phi_{f}|$).

Thus, from the above two cases, it is visible that the system's energy function has oscillatory response in phase-plane regions excluding phase error switching axis. Since the energy function based on phase error is changing only on $\Delta{\phi_{f}}$-axis vicinity, the Lyapunov function is evaluated only at phase error switching instant, as per Case I.

\subsection{Loop Gain constraint from Lyapunov Function}
In the DPLL under consideration, BBPD based NLTI subsystem is activated only when the phase error goes below a value that could be detected by the inverter based TDC. In CMOS65nm-LL technology, a single inverter delay of 20\,ps with reference clock of 100\,MHz, corresponds to the phase-error ($\phi_{err\_2}$) of 0.01 radians. For BBPD based DPLL activated in region $(|\phi_k|,|\Delta\phi_k|)<(0.01\,\mathrm{rad},0.01\,\mathrm{rad})$, the stability condition is derived for $(|\phi_k|_{max},|\Delta\phi_k|_{max})$ in Quadrants (I)/(III) of the phase-plane. As stated earlier, the state-space equations (\ref{eqn:phi_bbpd}) show that the system's trajectory moves in clockwise direction and the phase error sign reversal occurs only in the quadrants where $\phi_k$ and $\Delta\phi_k$ have similar sign. Thus, the constraint on loop-gain (${K'_{P3}},{K'_{I3}}$) is derived by replacing $(|\phi_k|_{max},|\Delta\phi_k|_{max})$ values in  (\ref{eqn:bbpd_energy_change}) with $(0.01\mathrm{rad},0.01\mathrm{rad})$ for this DPLL design, as shown in (\ref{eqn:phase_err_sign_reverse2}).

\begin{center}
$1001(2K_{P3}'+K_{I3}')^2-20.02(2K_{P3}'+K_{I3}')+0.0003<0$
\end{center}
\begin{equation}
\label{eqn:phase_err_sign_reverse2}
\Rightarrow0.00001\,\mathrm{rad}<(2K_{P3}'+K_{I3}')<0.02\,\mathrm{rad}
\end{equation}

The stability condition is required to be satisfied only outside the limit-cycle region i.e. $(\phi_k|_{min},|\Delta\phi_k|_{min}))\approx(2K_{P3}'+K_{I3}',2K_{P3}'+K_{I3}')$. Hence, the loop-filter parameters can be derived from (\ref{eqn:phase_err_sign_reverse2}) as per chosen Lyapunov candidate function. The loop gain values ($K_{P3},K_{I3}$) calculated from the linearized approximation of BBPD based DPLL in (\ref{eqn:gain_ratio}) also satisfies the negative derivative condition on Lyapunov function.  

The phase-portrait of BBPD based DPLL in Fig.\ref{fig:phase_portrait_bbpdpll} shows that with the choice of loop gain $K_P$/$K_I$, either settling time or jitter (from bang-bang operation) could be reduced at the expense of other. Figure \ref{fig:energy_fn_bbpdpll} shows a decreasing energy plot for this subsystem, with Lyapunov function being calculated only on phase-error sign-reversal boundary.

\begin{figure}[!h]
{
\subfloat[]{
\includegraphics[scale=0.17]{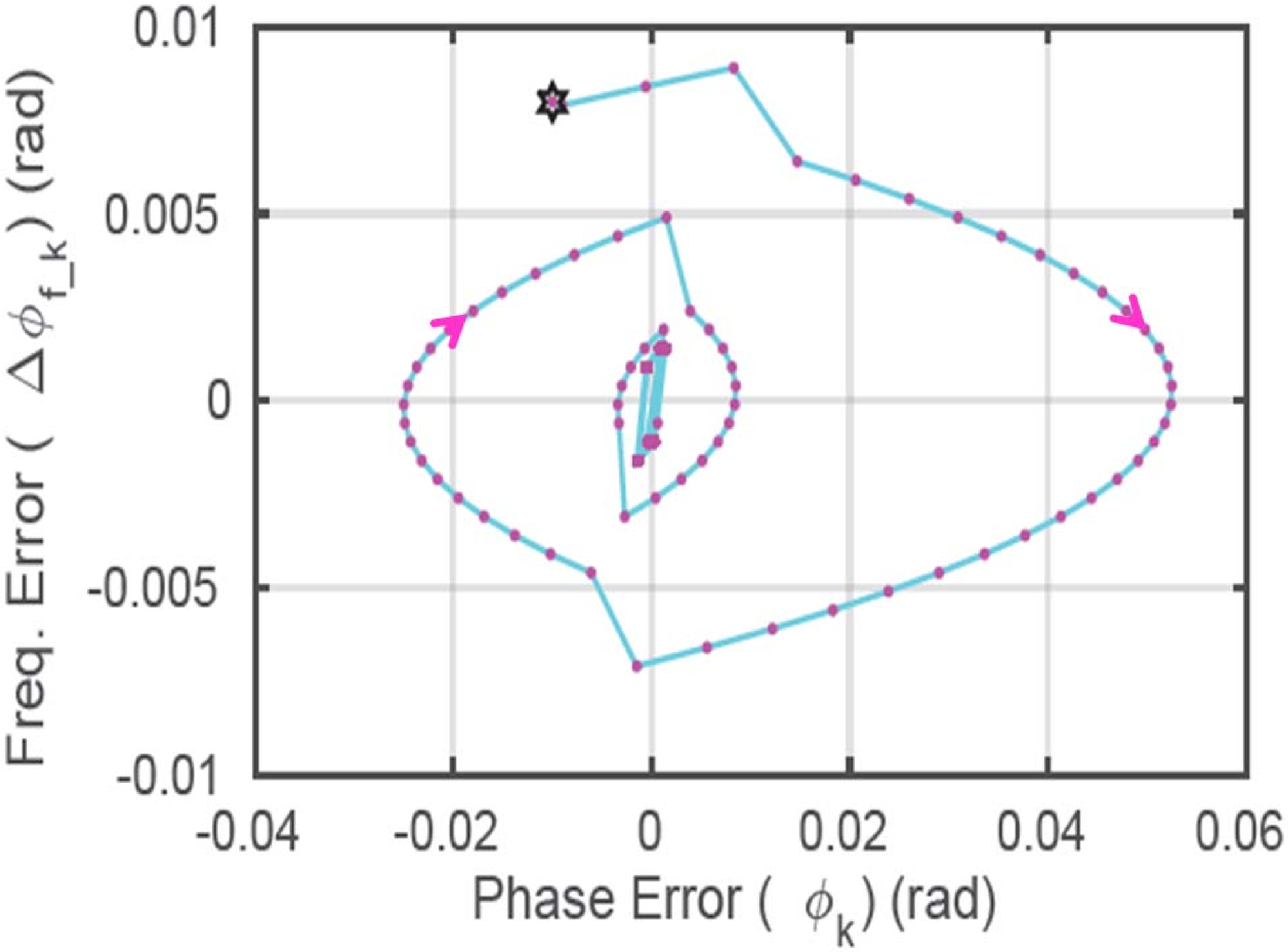}}
\subfloat[]{
\includegraphics[scale=0.16]{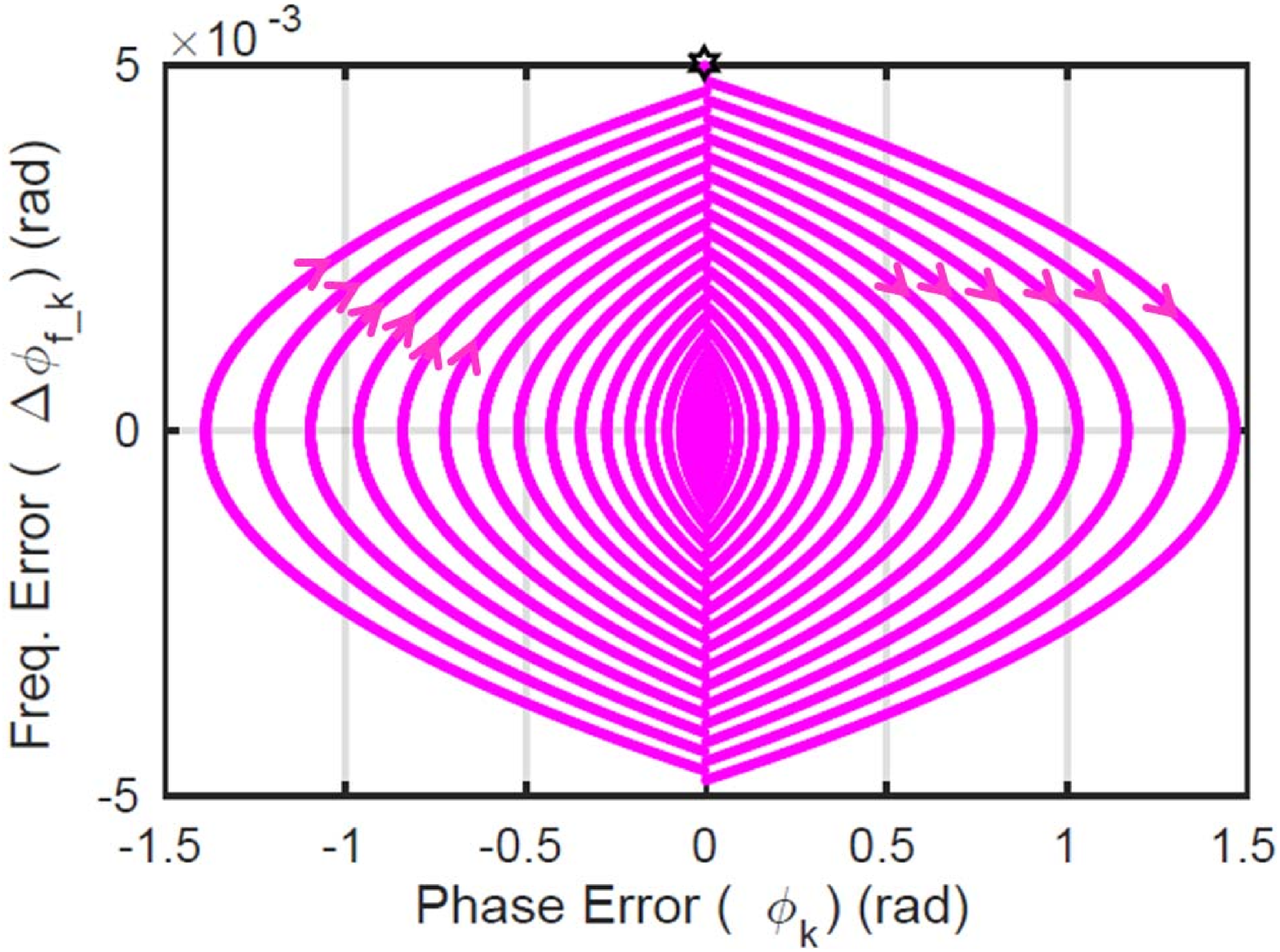}}
}
\caption{Phase-portrait of DPLL in BBPD mode with loop gains: (a) $K_{P3}'=0.001\,\mathrm{rad}$, $K_{I3}'=0.0005\,\mathrm{rad}$ (b) $K_{P3}'=0.00006\,\mathrm{rad}$, $K_{I3}'= 0.0000078\,\mathrm{rad}$.}
\label{fig:phase_portrait_bbpdpll}
\end{figure}

\begin{figure}[!h]
\centerline{\includegraphics[scale=0.32]{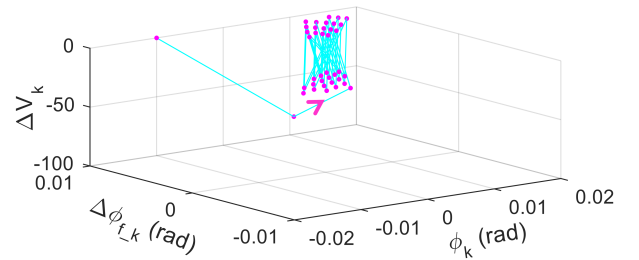}}
\caption{Energy decrease rate of DPLL in BBPD mode, calculated only at phase-error sign-reversal instants.}
\label{fig:energy_fn_bbpdpll}
\end{figure}

\subsection{Limit Cycle Stability}
Multiple Lyapunov  functions are associated  with  the transitions in  the  hybrid  system  so that the trajectory is shown to converge to the switch points of the limit cycle. The multiple QLF aims to focus on a local behavious of the system in each region. Construct a Lyapunov function outside the LaSalle's invariant set. This approach only evaluates the global stability of the system, while neglects the detailed behaviour within the invariant set.

There exist a stable limit cycle to which the continuous trajectory converge.

For a standalone BBPD based DPLL, the stability analysis in \cite{ref:nonlinear_analsys} suffices to derive the loop filter gain from (\ref{eqn:gain_ratio}). However, for BBPD based DPLL incorporated in a switched system, its Lyapunov function search is still required to prove overall system's stability using Multiple Lyapunov Functions approach.

\section{BBPD+FSM based DPLL stability}
\label{sec:fsm_stability}
During the locking process in the considered DPLL, when the phase error ($|\phi_{k}|$) reduces below $\phi_{err\_2}$($\approx 0.01\,rad$) boundary, the system switches from LTI-subsystem to BBPD+FSM mode. With the activation of FSM, the derivative gain of Differentiator state is initialized with a  value 
($K_{D,init}$) required for the remaining phase error correction. For an immediate phase correction within bounds, the derivative gain is decremented ($K_{D,k}/\beta$) by FSM-Differentiator state, during each phase error sign-reversal. In this FSM, an additional accumulator ($K_{I,fsm}$) is activated for fast frequency tracking  when consecutive cycles have same phase-error sign. The system switches from BBPD+FSM mode to BBPD based NLTI subsystem, when the derivative gain ($K_{D,k}$) is reduced to a value of 1. The deactivation of FSM ensures that the limit-cycle region of DPLL doesn't extend beyond the area due to BBPD operation, thus improving the PLL jitter performance. The state-space equation for FSM-Integrator stage and FSM-Differentiator stage is described in (\ref{eqn:integ_state_space}) and (\ref{eqn:diff_state_space}) respectively. 
\begin{equation}
\label{eqn:integ_state_space}
\begin{aligned}
&\phi_{k+1}=\phi_{k}+\Delta{\phi_{f\_k}} - K_{P3}'\sigma(\phi_{k}) - K_{I3}'K_{I,fsm_{k}}\sigma(\phi_{k}) \\
&\Delta\phi_{f\_k+1}=\Delta{\phi_{f\_k}} - K_{P3}'\sigma(\phi_{k}) - K_{I3}'K_{I,fsm_{k}}\sigma(\phi_{k}) \\
&K_{I,fsm_{k+1}}=K_{I,fsm_{k}}+1
\end{aligned}
\end{equation}
\begin{equation}
\label{eqn:diff_state_space}
\begin{aligned}
&\phi_{k+1}=\phi_{k}+\Delta{\phi_{f\_k}} - K_{D,k}(K_{P3}'+K_{I3}')\sigma(\phi_{k}) \\
&\Delta\phi_{f\_k+1}=\Delta{\phi_{f\_k}} - K_{D,k}(K_{P3}'+K_{I3}')\sigma(\phi_{k}) \\
\end{aligned}
\end{equation}

Figure \ref{fig:fsm_trajectory}(a) shows the diverging trajectory of FSM-Integrator based DPLL governed by its state-space equations. Figure \ref{fig:fsm_trajectory}(b) shows the stability of  FSM-Differentiator state with its negative energy-derivative rate calculated as per Lyapunov function defined in (\ref{eqn:bbpd_energy}).

\begin{figure}[!h]
{
\subfloat[]{
\includegraphics[scale=0.15]{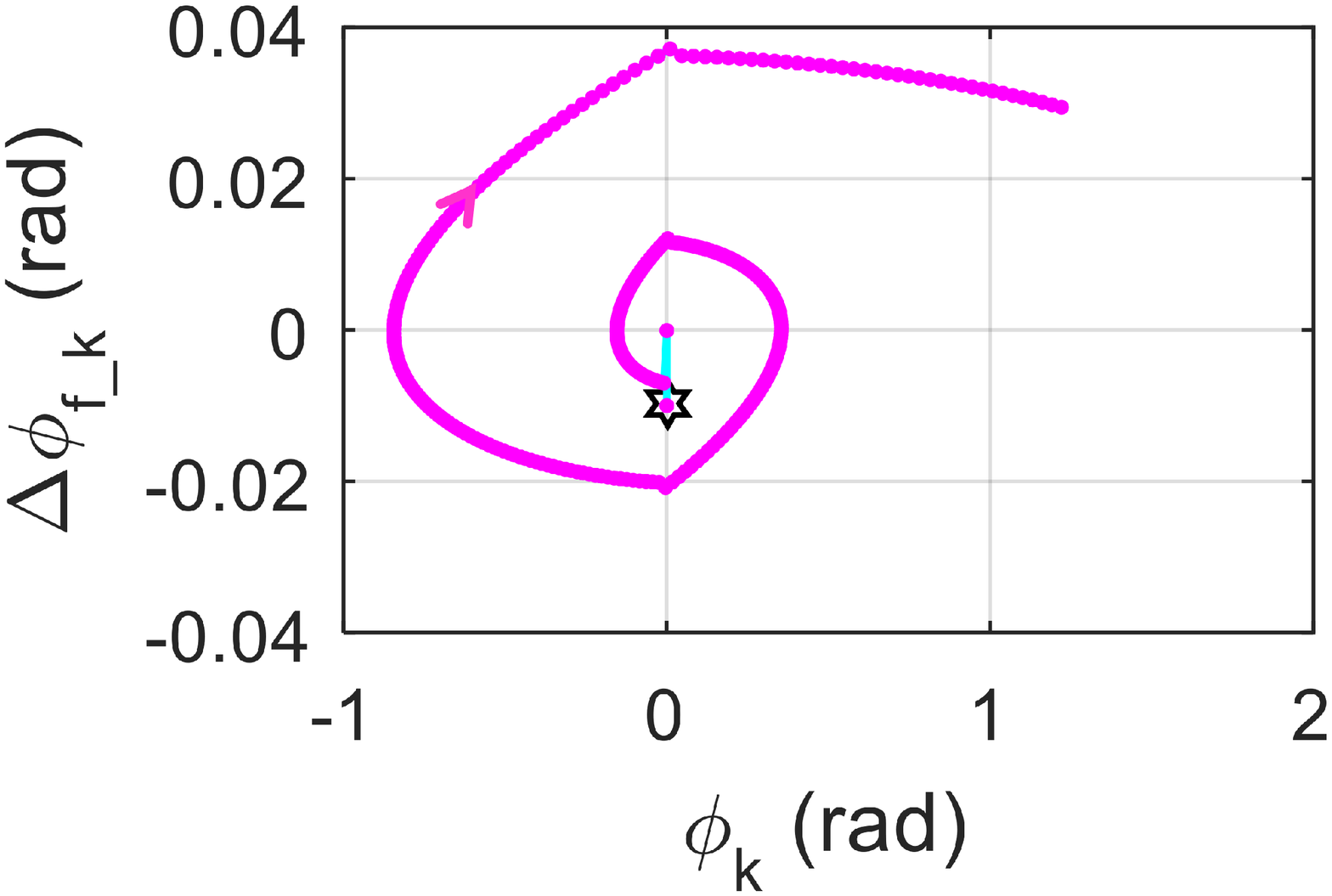}}
\subfloat[]{
\includegraphics[scale=0.19]{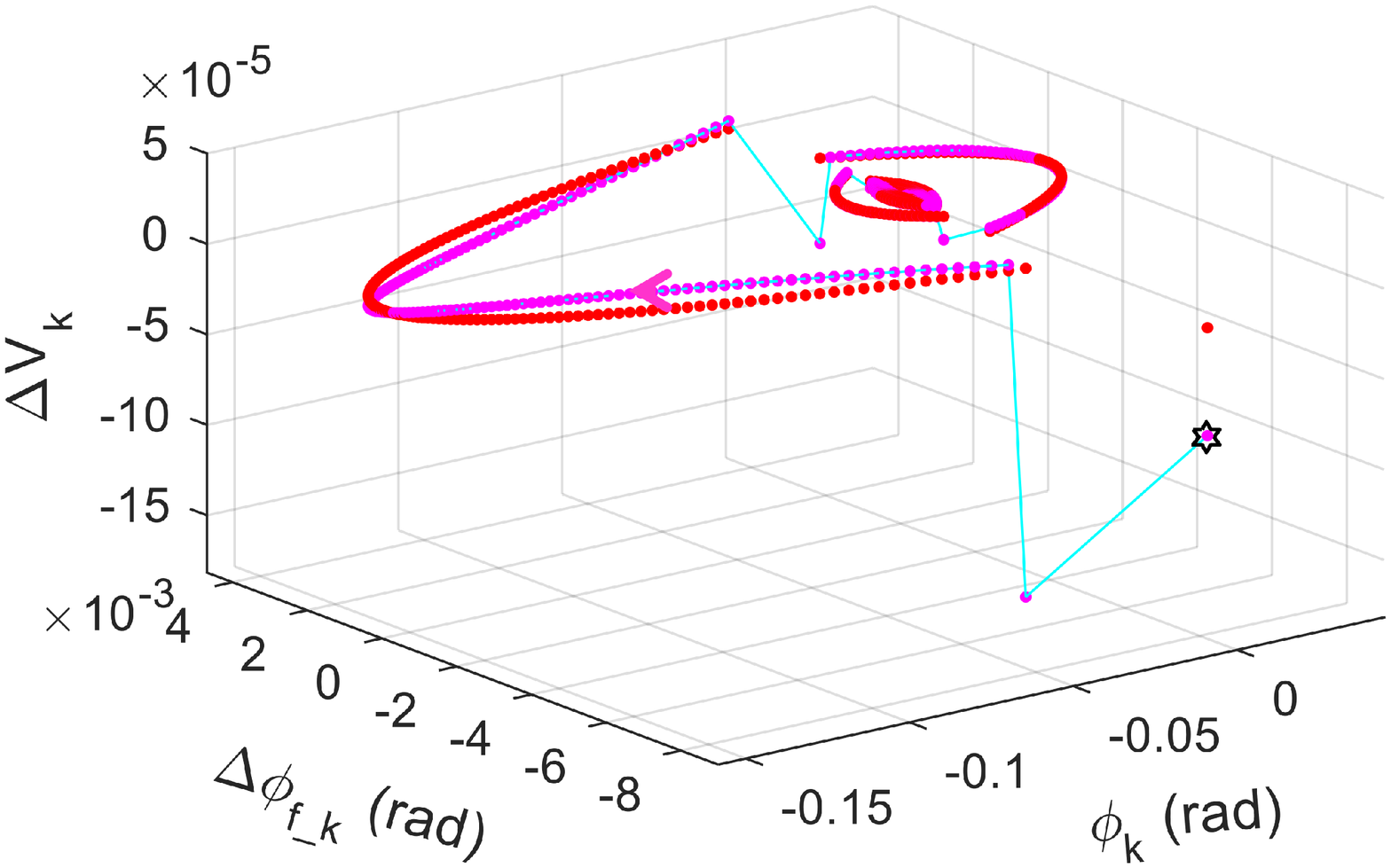}}
}
\caption{(a) Diverging trajectory for FSM-Integrator state; (b) Negative energy-derivative for FSM-Differentiator state with $K_{D,init}=64$, $K_{P3}'=0.00006\,\mathrm{rad}$, $K_{I3}'= 0.0000078\,\mathrm{rad}$ and derivative gain reduction by half at every phase-error sign-reversal.  }
\label{fig:fsm_trajectory}
\end{figure}

The presence of an additional integrator in the loop, as one of the FSM states, makes the system unstable during its activation period. Therefore, it is important to analyze the loop under this brief instability, with a fixed switching law defined by the FSM. With the FSM controlling the DPLL, unstable integrator state is either followed by LTI-subsystem activation (for $|\phi_{k}|>0.01rad$) or with Differentiator-state activation (for $\phi_{k}$ sign-reversal). Accordingly, for stability analysis, Lyapunov function ($V_i$) is derived for (i) Integrator+LTI subsystem and (ii) Integrator+Differentiator subsystem, only in the suitable region of activation of each state. The overall system can then shown to be asymptotically stabilizable, if $V_j$ at the beginning of each interval on which the $j^{th}$ subsystem is active is not exceeding the value at the beginning of the previous such interval.

As discussed in Section \ref{sec:bbpd_stability}, the system's trajectory described by (\ref{eqn:integ_state_space}) moves in clockwise direction on the phase-plane. Thus, Integrator+LTI subsystem is evaluated in Quadrants I/III of phase-plane where $|\phi_{k}|$ increases, and Integrator+Differentiator subsystem is activated in Quadrants II/IV where $|\phi_{k}|$ decreases. Figure \ref{fig:fsm_phase_plane} shows the initial-point and end-point range along with the region of activation of each subsystem. 

\begin{figure}[!h]
\centerline{\includegraphics[scale=0.65]{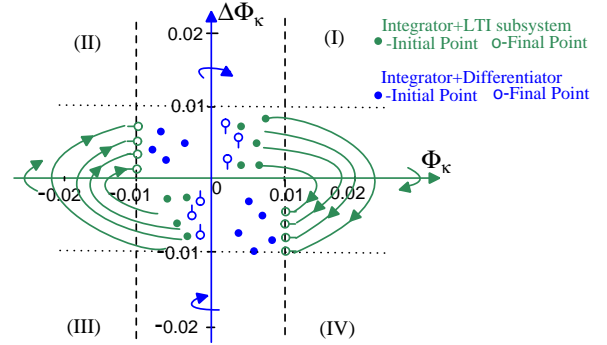}}
\caption{BBPD+FSM based DPLL phase plane divided into regions where (i) Integrator+LTI subsystem is active, or (ii) Integrator+Differentiator is active.}
\label{fig:fsm_phase_plane}
\end{figure}

Following section discusses stability analysis of FSM-states for region having \{$|\phi_k|,|\Delta\phi_{f\_k}|$\}$<$\{$0.01\,\mathrm{rad},0.01\,\mathrm{rad}$\}, because beyond this region, correction ($K_{P3}',K_{I3}'$) offered by the FSM-enabled subsystems is negligible in comparison to $(|\phi_{k}|,|\Delta\phi_{f\_k}|)$. Across this region, the loop trajectory is mainly governed by state-space equations of LTI subsystem. 

\subsection{Integrator+LTI subsystem stability}
In Quadrants I-III, starting from any arbitrary point, system's trajectory follows Integrator state-space equations defined in (\ref{eqn:integ_state_space}) for $|\phi_{k}|<0.01\,rad$, and LTI state-space equations defined in (\ref{eqn:phi_lin2}) for $|\phi_{k}|>0.01\,rad$.  Based on the filter gain derived from  Sections \ref{sec:lti_stability_analysis}-\ref{sec:bbpd_stability}, the system trajectory could be predicted. Figure \ref{fig:fsm_region} shows that for $|\Delta\phi_{init}|<0.001\,rad$, the loop does not switch  to LTI-mode as phase error remains restricted  below $|\phi_{err\_2}|$. With the chosen filter gain values as in Table \ref{tab:lyapunov_condition}, the range of initial and terminating points for this subsystem's trajectory are :
\begin{center}
 $|\phi_{init}|_{min}\approx0\,rad, |\phi_{final}|_{max}=0.01\,rad,$ \\ $0.001\,rad<|\Delta\phi_{init}|<0.01\,rad$
\end{center}.

\begin{figure}[h]
\centerline{\includegraphics[scale=0.25,trim=0.4cm 0.2cm 0.4cm 0 , clip]{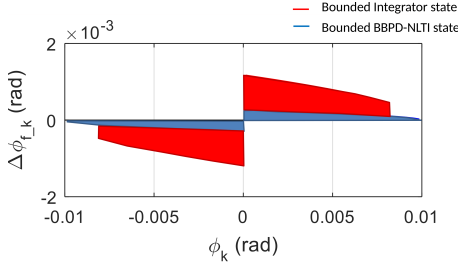}}
\caption{DPLL phase-plane regions with trajectory governed by (a) FSM-integrator state-space equations or (b) BBPD based DPLL state-space equations, wherein starting from any arbitrary point, the phase-error remains bounded within 0.01\,rad, thus, terminating any further switching out of BBPD+FSM subsystem. }
\label{fig:fsm_region}
\end{figure}

Consider the energy function $V_{1,k}$ defined by (\ref{eqn:cqlf}) at switching-in point ($\phi_{init},\Delta\phi_{init}$) and switching-out point ($\phi_{final},\Delta\phi_{final}$) of this subsystem:
\begin{center}
 $V_{1,k}=0.02{\phi_k^2}+0.12\phi_k\Delta\phi_{f\_k}+3{\Delta\phi_{f\_k}^2}$
\end{center}

For $0.001\,rad<|\Delta\phi_{init}|<0.01\,rad$, the system's trajectory based on (\ref{eqn:integ_state_space})(\ref{eqn:phi_lin2}) converges to $|\Delta\phi_{final}|<0.002\,rad$, as shown in Fig. \ref{fig:intg+lti2}(a). Hence, for any arbitrary point in the above defined range, $n^{th}$-derivative of Lyapunov function $\Delta{V_{1,(k+n)-k}}$ calculated as the energy difference at initial and terminating point of the subsystem is negative. This shows that energy increment in the loop caused by unstable integrator state is overcome by the large gain of LTI-subsystem. The energy decrement pattern across trajectory defined by Integrator+LTI subsystem is shown in Fig \ref{fig:intg+lti2}(b). 

\begin{figure}[!h]
{
\subfloat[]{
\includegraphics[scale=0.18,trim=1cm 0 1.5cm 0, clip]{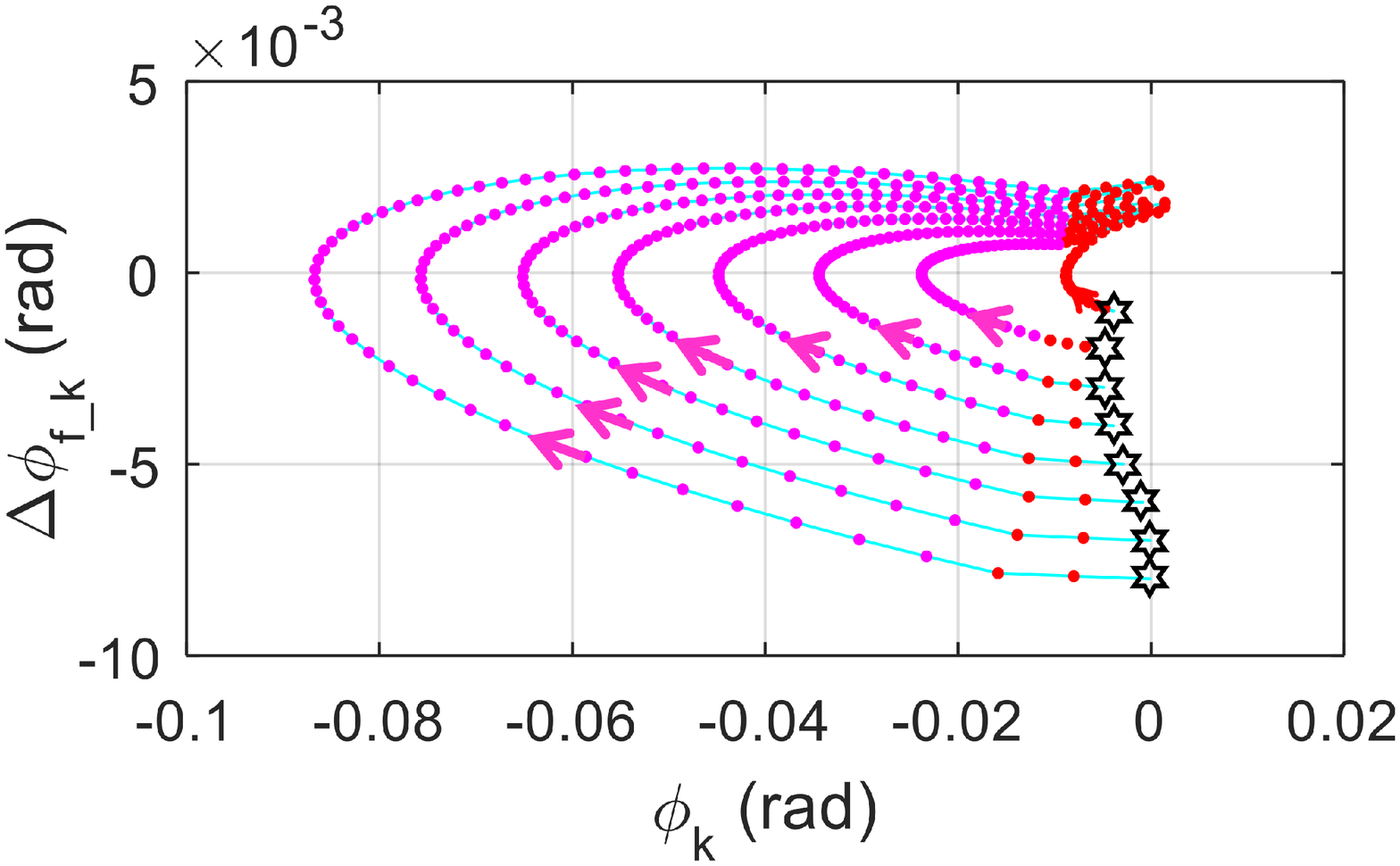}}
\subfloat[]{
\includegraphics[scale=0.18]{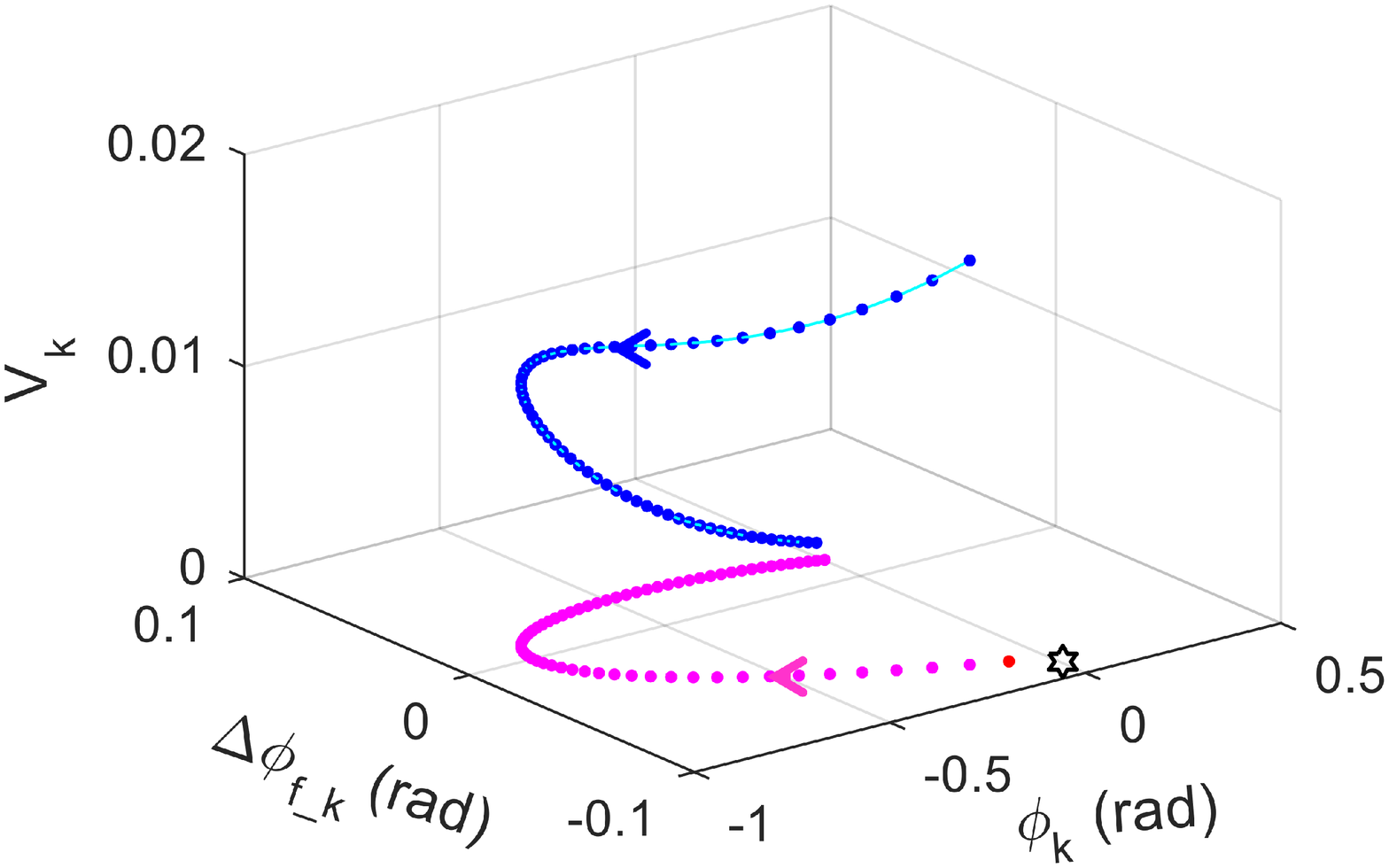}}
}
\caption{(a) DPLL trajectory governed by Integrator+LTI state-space equations; (b) Energy plot for DPLL behaviour governed by Integrator+LTI subsystem.}
\label{fig:intg+lti2}
\end{figure}


\subsection{Integrator+Differentiator subsystem stability}
With the phase-error decrease in Quadrants II-IV, the system's trajectory follows Integrator state-space equation (\ref{eqn:integ_state_space}) for $0\,\mathrm{rad}<|\phi_{err}|<0.01\,\mathrm{rad}$, and Differentiator state-space equation defined by (\ref{eqn:diff_state_space}) on phase-error sign-reversal.
Since BBPD+FSM mode is activated in the DPLL only for $|\phi_{err}|<0.01\,\mathrm{rad}$, the initial derivative gain ($K_{d,init}$) stability is evaluated in this region.  As per (\ref{eqn:diff_state_space}), FSM-Derivative state gives a constant correction of $K_{D,init}(K_{P3}'+K_{I3}')$ to the loop irrespective of whether the current point in state-space is near the equilibrium region or away from it. In case, the PLL is already near the locked region, a large derivative gain will increase the energy in the system instead of reducing it. Figure \ref{fig:intg+lti2} highlight that the Integrator+LTI subsystem is able to reduce phase-error to a magnitude of 0.002\,rad in the implemented DPLL. For providing further correction, the derivative gain is decided with condition as 
\begin{equation}
\label{eqn:kd_init1}
0.002\,\mathrm{rad}<K_{D,init}(K_{P3}'+K_{I3}')<0.01\,rad.
\end{equation} 

\noindent For FSM based DPLL to have settling response analogous to binary-search algorithm, $K_{D,init}$ is chosen as mid-point of the range defined by (\ref{eqn:kd_init1}) i.e. [{$K_{D,init}(K_{P3}'+K_{I3}')\approx0.005 rad$}]. 

While BBPD+FSM mode is active, this derivative gain is reduced by a factor of $\beta$ at each phase error sign-reversal instant to reduce the limit-cycle region in locked state. The derivative gain reduction factor is chosen as 2 in this design, so that the loop trajectory could be eventually placed in the bounded region of FSM-Integrator state, as shown in Fig. \ref{fig:fsm_region} and Fig.\ref{fig:intg+diff}(a). Within this bounded region of Integrator state, the phase error remains below $0.01\,\mathrm{rad}$ which ensures that the system doesn't switch back to LTI-mode again. Figure \ref{fig:intg+diff} shows that with chosen $K_{D,init}$, the phase or frequency error decreases at switching-out point in comparison to the switching-in point of this subsystem, with Lyapunov function being 

\begin{equation}
 V_{1,k}=0.02{\phi_k^2}+0.12\phi_k\Delta\phi_{f\_k}+3{\Delta\phi_{f\_k}^2}.
\label{eqn:fsm_deriv_energy}
\end{equation}

\begin{figure}[!h]
\centering{
\includegraphics[scale=0.23,trim=0.4cm 0 2cm 0, clip]{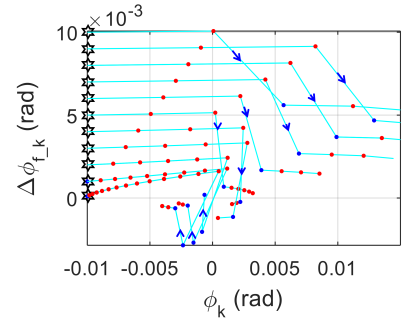}
}
\caption{DPLL trajectory governed by Integrator+Derivative state-space equations with $K_{D,init}=64, K_{P3}'=0.00006\,\mathrm{rad}, K_{I3}'= 0.0000078\,\mathrm{rad}, \beta=2$.}
\label{fig:intg+diff}
\end{figure}
\begin{table*}[!h]
 \renewcommand{\arraystretch}{1.4}
 \caption{Lyapunov Functions for DPLL subsystems mode}
\label{tab:lyapunov_condition}
\begin{center}
\renewcommand{\arraystretch}{1.4}
  \begin{tabular}{  l  c c c}
    \hline
\textbf{DPLL} \textbf{Subsystems} &  \textbf{Lyapunov Functions} & \textbf{Activation}  \textbf{Region} & \textbf{Loop Filter Gain} \\ \hline
LTI-1  & $V_{1,k}=0.02{\phi_k^2}+0.12\phi_k\Delta\phi_{f\_k}+3{\Delta\phi_{f\_k}^2}$ & $\phi_{err} > 1 \mathrm{rad}$ &  $K_{P1}'=0.03\mathrm{rad}$, $K_{I1}'=0.007\mathrm{rad}$\\ 
LTI-2  & $V_{1,k}=0.02{\phi_k^2}+0.12\phi_k\Delta\phi_{f\_k}+3{\Delta\phi_{f\_k}^2}$ & $1 \mathrm{rad} > \phi_{err} > 0.01 \mathrm{rad}$ & $K_{P2}'=0.05\,rad$, $K_{I2}'=0.003\,rad$\\ \hline
Integrator+LTI-2  & $V_{1,k}=0.02{\phi_k^2}+0.12\phi_k\Delta\phi_{f\_k}+3{\Delta\phi_{f\_k}^2}$ & $\phi_{err} < 0.01 \mathrm{rad}$ & \\
&    (Evaluated at switching-in and switching-out point)  & ($K_D > 0$, Quadrant I/III) & \\                                                                 
Integrator+Differentiator& $V_{1,k}=0.02{\phi_k^2}+0.12\phi_k\Delta\phi_{f\_k}+3{\Delta\phi_{f\_k}^2}$ &  $\phi_{err} < 0.01 \mathrm{rad}$ & $K_{D,init} = 64$\\
 & (Evaluated at switching-in and switching-out point)& ($K_D > 0$, Quadrant II/IV) &\\ 
BBPD based NLTI & $V_{3,k}= {\phi_k^2}+1000{\Delta\phi_{f\_k}^2}$ & $\phi_{err} < 0.01\,\mathrm{rad}$ & $K_{P_3}'=0.00006\,\mathrm{rad}$\\
     & (Evaluated outside limit-cycle region) & $(K_D = 0)$ & $K_{I_3}'=0.0000078\,\mathrm{rad}$\\ \hline
    \end{tabular}
\end{center}
\end{table*}

\section{DPLL stability as switched system}
\label{sec:sys_stability}
Figure \ref{fig:phase_err_fsm} shows that based on the magnitude of the phase-error ($|\phi_{err}|$), the system switches between PFD based LTI-mode, BBPD+FSM based NLTV-mode or standalone-BBPD based NLTI mode. Once the phase-error decreases below $|\phi_{err2}|$ boundary, the BBPD+FSM is activated introducing either Integrator+LTI subsystem or Integrator+Differentiator subsystem, based on the quadrant where current-state of system is positioned. Table \ref{tab:lyapunov_condition} highlights that PFD based LTI-systems with an adaptive loop bandwidth and BBPD based FSM could be designed with common Lyapunov function ($V_{1,k}$), thus assuring stability for arbitrary switching between these subsystems. The stability condition for BBPD based DPLL is derived with Lyapunov function ($V_{3,k}$) defined in (\ref{eqn:bbpd_energy}). The Integrator+Differentiator state ensures that the DPLL is placed in limit-cycle region, thus avoiding chattering phenomenon out of BBPD mode. Figure \ref{fig:dpll_stability} shows the trajectory convergence of switched-DPLL system while traversing through phase-plane regions governed by different subsystems.

\begin{figure}[h]
\centerline{\includegraphics[scale=0.3]{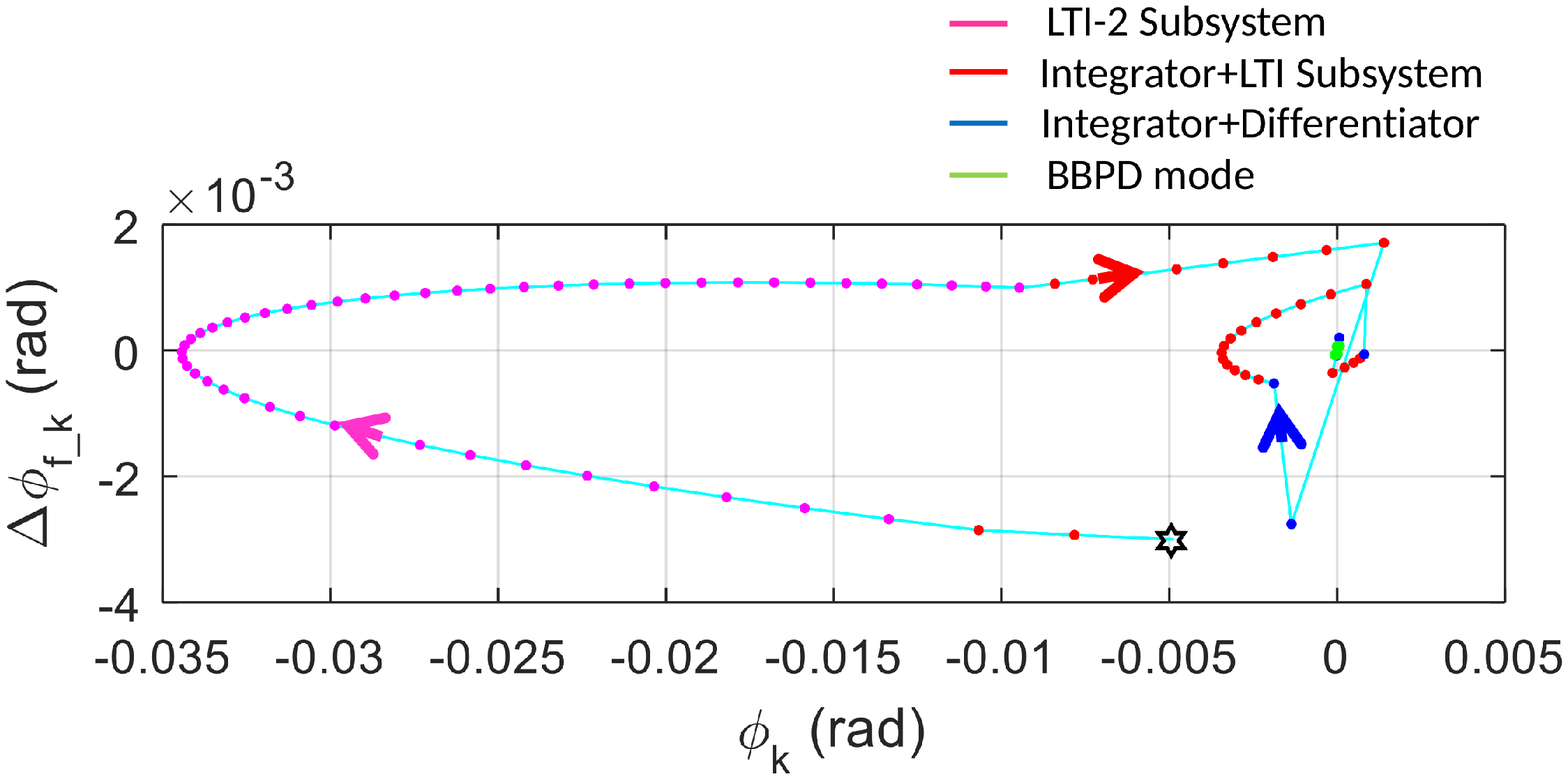}}
\centerline{\includegraphics[scale=0.3]{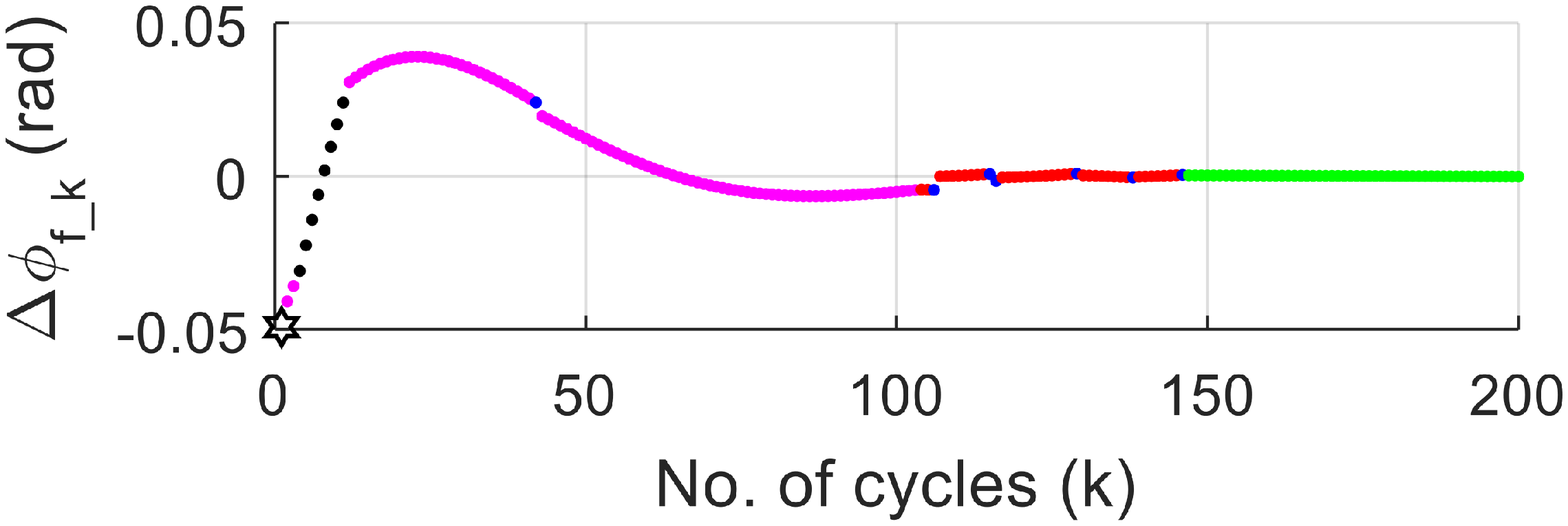}}
\caption{Converging trajectory of switched-DPLL system.  }
\label{fig:dpll_stability}
\end{figure}


\section{Measurement Results}
\label{sec:meas_result}
The switched DPLL system, discussed in this work, is implemented in CMOS65nm-LL technology with chip micrograph as shown in Fig \ref{fig:micrograph}. Figure \ref{fig:dpll_settling_time} highlights the settling response of the DPLL with loop gain parameters derived based on the stability analysis discussed in this work. Figure \ref{fig:dpll_jitter} shows that the DPLL achieves lock without chattering between different subsystems, thus constraining the output jitter only by the limit cycle region of BBPD based NLTI mode.

\begin{figure}[h]
\centering{\includegraphics[scale=0.5]{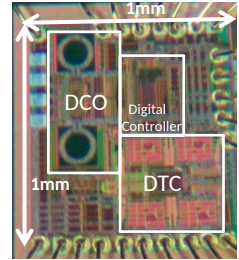}}
\caption{Chip Micrograph of DPLL.  }
\label{fig:micrograph}
\end{figure}

\begin{figure}[!h]
\centering{\includegraphics[scale=0.205]{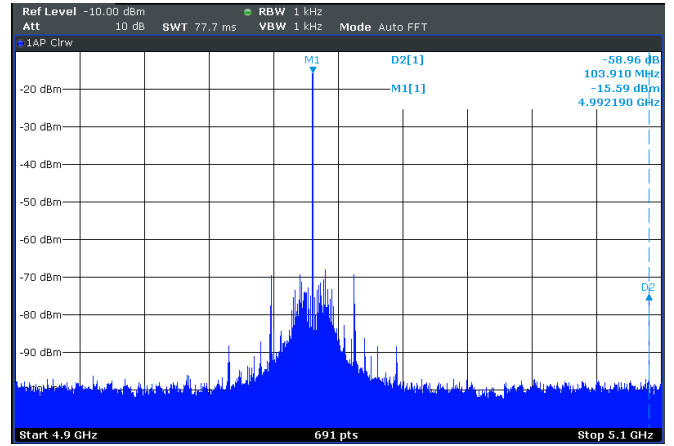}}
\caption{Measured spectrum of 5\,GHz DPLL with reference spur rejection of 59\,dB.} 
\label{fig:dpll_settling_time}
\end{figure}

\begin{figure}[!h]
\centering{\includegraphics[scale=0.35]{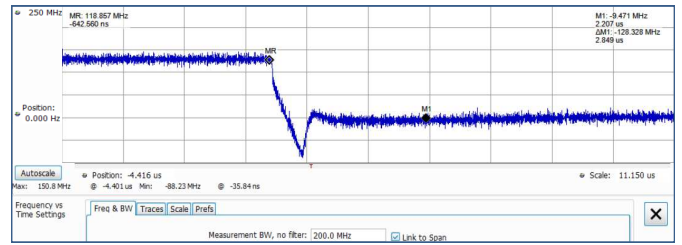}}
\caption{Measured settling response of 5\,GHz fractional-N DPLL for 128\,MHz frequency step change.} 
\label{fig:dpll_settling_time}
\end{figure}

\begin{figure}[!h]
\centering{\includegraphics[scale=0.31]{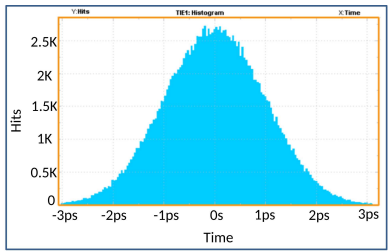}}
\caption{Measured jitter histogram of fractional-N DPLL.} 
\label{fig:dpll_jitter}
\end{figure}

Figure \ref{fig:new_fom} highlights a competitive Figure of Merit ($FoM_{t_s}$) \cite{ref:dpll_fom} and best  lock time being achieved by the fractional-N DPLL incorporating the switched loop technique in the feedforward path. The Figure of Merit used for DPLL performance benchmarking is given as
\begin{equation}
\label{eqn:fom_adpll}
\mathrm{FoM} = 10\mathrm{log}\left[\left(\frac{\sigma_t}{1\mathrm{s}}\right)^2\left(\frac{t_s}{1\mathrm{s}}\right)^2\left(\frac{P}{1\mathrm{mW}}\right)\right],
\end{equation}

\noindent where ${\sigma_t}$ is the output jitter, ${t_s}$ is the lock time and $P$ is the power consumptionof the DPLL.
\begin{figure}[!h]
\centering
	\includegraphics[scale = 0.53]{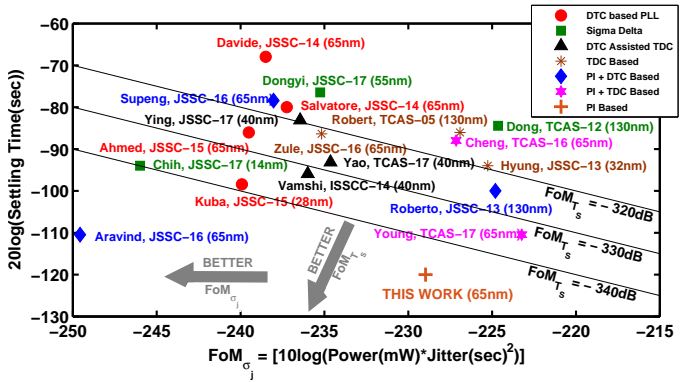}
\caption{Fractional-N DPLL performance benchmarking based on jitter, power and settling time response.}
\label{fig:new_fom}
\end{figure}

\section{Conclusion}
This work highlights the stability analysis using Lyapunov function towards deriving the loop parameters for a switched DPLL system. The DPLL phase plane is partitioned into regions based on the phase-error state dependent switching across different subsystems. Multiple Lyapunov functions are used towards deriving the stability conditions for these subsystems. The measured fast settling response of the DPLL implemented with derived loop parameters, proves that the system achieves phase-lock without chattering phenomenon between different subsystems. The illustrated stability analysis for the DPLL unfurls the possibility of engaging an unstable integrator or an impulsive differentiator, without risking the convergence of the system. 

The discussed stability analysis  only verifies the loop convergence for a predefined switching rule and loop gain values obtained from first-order approximations. As a future scope to this work, an analysis could be developed to derive an optimum state-dependent switching rule and loop gain values for attaining maximum performance from the loop-order switching in a DPLL.


\bibliographystyle{IEEEtran}
\bibliography{ref}

\end{document}